\documentclass[A4paper]{extended-abstract-2023}

\usepackage{graphicx}
\usepackage{amsmath}
\usepackage{amsfonts}
\usepackage{amssymb}
\usepackage[symbol]{footmisc}

\usepackage[hidelinks]{hyperref} 
\usepackage[a4paper,
left=0.8in,
right=0.8in,
top=1.2in,
bottom=0.5in,]{geometry}

\usepackage{multirow}
\usepackage{amsmath}
\usepackage{mathtools}
\usepackage{tikz}
\usepackage{booktabs} 
\usepackage{tabularx}
\newcolumntype{C}[1]{>{\centering\arraybackslash}p{#1}} 
\newcolumntype{L}[1]{>{\raggedright\arraybackslash}p{#1}} 
\newcolumntype{R}[1]{>{\raggedleft\arraybackslash}p{#1}} 

\usepackage[utf8]{inputenc}

\usepackage{acronym}
\acrodef{FEM}{finite element method}
\acrodef{MOR}{model order reduction}
\acrodef{fROM}{frequency-domain reduced order model}
\acrodef{rA-Krylov}{rational Arnoldi Krylov subspace method}
\acrodef{SPL}{sound pressure level}
\acrodef{CFRP}{carbon-fibre-reinforced plastic}
\acrodef{DoF}{degrees of freedom}
\acrodef{FRF}{frequency response functions}
\acrodef{DD}{domain decomposition}

\usepackage{color}
\definecolor{ina}{rgb}{0.67, 0.76, 0.23}

\usepackage[colorinlistoftodos,prependcaption,textsize=scriptsize]{todonotes}

\usepackage{pgfplots}
\pgfplotsset{compat=newest}
\usepgfplotslibrary{groupplots}
\usepgfplotslibrary{dateplot}

\usepgfplotslibrary{colormaps}
\pgfplotsset{
	colormap={warm}{
		rgb255=(121,23,23)
		rgb255=(181,1,1)
		rgb255=(239,71,25)
		rgb255=(249,131,36)
		rgb255=(255,180,0)
		rgb255=(255,229,6)
	},
}

\usepackage{subcaption}
\usepackage[numbers,sort&compress]{natbib}
\title{Efficient solution strategies for cabin noise assessment of a wave resolving aircraft fuselage model}

\author{Christopher Blech\footnote{Corresponding author: Christopher Blech, c.blech@tu-braunschweig.de}, Harikrishnan K. Sreekumar, Yannik Hüpel,\\ Sabine C. Langer}

\heading{C. Blech, H. K. Sreekumar, Y. Hüpel and S. C. Langer}

\address{Institute for Acoustics and Dynamics, Technische Universität Braunschweig,\\ Braunschweig, Germany
\and
Cluster of Excellence SE$^2$A–Sustainable and Energy-Efficient Aviation,\\Technische Universität Braunschweig, Germany}

\abstract{For the purpose of high-fidelity aircraft cabin noise simulations during early design phases, we study three efficient solving approaches for the fully coupled finite element model of an aircraft fuselage segment. Obtaining an efficient solution with respect to consumed computational time and resources is challenging within a conventional simulation pipeline, as large-scale and complex vibroacoustic models demand crucially high computational costs with increasing frequency. In this contribution, we adopt (1) frequency and domain-adaptive discretisation, (2) domain-decomposition techniques, and (3) model order reduction with rational Arnoldi Krylov subspace methods for an aircraft fuselage model. The three approaches have shown remarkable advantage thereby reducing the solving time as well as the memory requirement that are essential when solving large-scale models. While the discretisation and the model order reduction approaches accelerate the solving process by efficiently handling the complexity of the system to be solved, domain-decomposition techniques further handle the aspect of reducing the overall memory consumption. Finally with the help of active research aircraft models, we implement and showcase the achieved efficiency.}


\begin{document}
\thispagestyle{empty}

\section{INTRODUCTION}
\label{sec:Intro}
More and more people are exposed to cabin noise during flights as there is an increasing need for air connections of a growing population within an increasingly networked world. 
A key acceptance factor in the development of future aircraft is environmental friendliness, which consequently comprises low sound pressure levels within the cabin. 
For the purpose of an advantageous acoustic design, wave-resolving vibroacoustic models can be applied in early design stages \cite{langer2019cabin, peiffer2016full}. For example, novel aircraft concepts can be studied without many mechanical assumptions under realistic load distributions \cite{blech20jsv, blech20springer}.
This way, reasonable sound reduction measures can be found \cite{blech_internoise_2017,rothe_ica_22} and disturbing and harmful sound pressure levels avoided exploiting the full design potential. 
Wave-resolving models are commonly solved by discretisation methods such as the \ac{FEM}. 
Above a certain frequency, the numerical solution by direct or iterative approaches is generally challenging as the system matrices increase exponentially with decreasing wavelengths. The \textit{problem of short wave lengths} \cite{zienkiewicz2000achievements} is already mentioned and still visible today. Though computing capacities increased over decades, the maximum solvable frequency for large systems like aircraft still does not meet the maximum frequency of interest. Peiffer \cite{peiffer2016full} depicts an application of wave-resolving models of aircraft segments up to approximately $200\,$Hz. 
Furthermore, the underlying complex mechanical formulations yield badly conditioned system matrices and strong couplings between structural and acoustic domains comprising dependencies on frequency, which also makes the accessibility more difficult and plays an enormously important role in solution efficiency. 

In this paper, we consider a vibroacoustic FE model of an aircraft fuselage as a reference example for cabin noise calculations during aircraft design. Based on a novel all-electric regional aircraft concept within the cluster of excellence ``Sustainable and Energy Efficient Aviation'' at TU Braunschweig \cite{karpuk21}, an airframe design following a composite construction trimmed by typical insulation and interior linings is derived. The actual mechanical problem statement as well as reference time and memory solution efforts are introduced in Sec.~\ref{sec:problem}. On the basis of this model, we apply three promising solution strategies, namely domain-adaptive and frequency-dependent discretisations, \ac{DD} techniques and \ac{MOR} with \ac{rA-Krylov} in Sec.~\ref{sec:strategies}. For each approach, the benefits in efficiency are exploited and raised to a general level for cabin noise calculations. 
Especially during design, extensive parameter studies are helpful in order to derive modifications based on parameter sensitivities. Hence, a highly efficient solution of the model is indispensable for a reasonable and successful application of numerical studies. The paper is closed with a summary and recommendations on efficient solution strategies in early design stages within Sec.~\ref{sec:summary}.

\section{PROBLEM STATEMENT AND REFERENCE SOLUTION STRATEGY}
\label{sec:problem}

\begin{figure}[!htb]
	\centering
	{%
		\setlength{\fboxsep}{0pt}%
		\setlength{\fboxrule}{0pt}%
		\fbox{\input{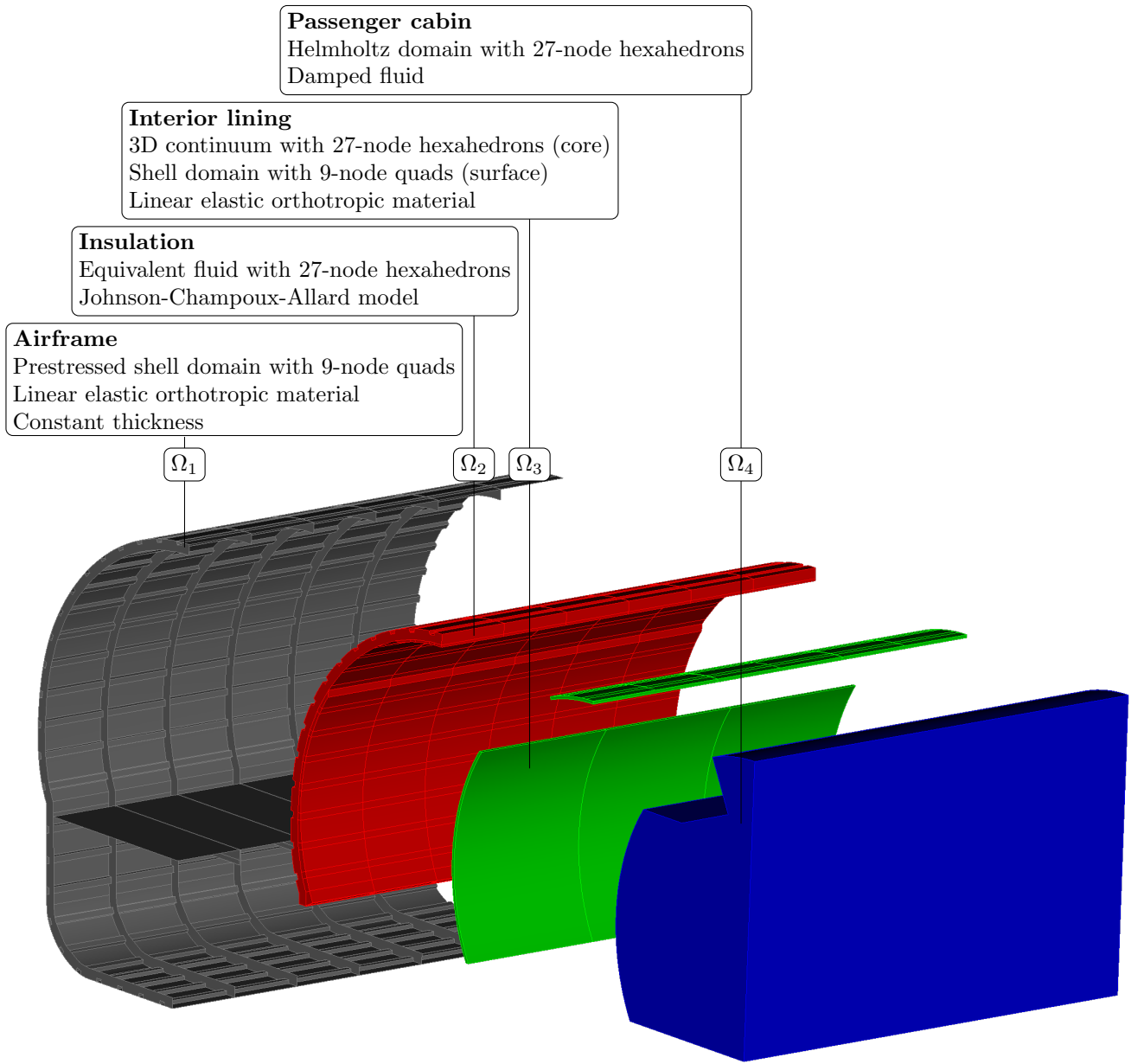}}}
	\caption{Vibroacoustic FE reference model comprising $2.43\,$mio DoFs within four major domains}
	\label{fig:modelOverview}
\end{figure}

For the vibroacoustic model, a fuselage segment of $3.0\,$m length ($5$ seat rows) is considered in order to calculate the \ac{SPL} in the cabin based on a generic plane wave excitation for the purpose of the underlying efficiency studies. The chosen length is assumed to be representative for recommendations on solution strategies as further extensions of a fuselage scale approximately linearly with regard to computational costs. As shown in Fig.~\ref{fig:modelOverview}, the model comprises four strongly coupled domains, namely the airframe $\Omega_1$, the insulation $\Omega_2$, the interior lining $\Omega_3$ and the cabin itself $\Omega_4$, which is finally the domain of interest. Generally, second-order standard finite elements with 9 or 27 nodes are used for the 2D and 3D domains, respectively, as described in Fig.~\ref{fig:modelOverview}. 

The \textbf{airframe} $\mathbf{\Omega_1}$ is made of \ac{CFRP} comprising an outer skin and a floor as well as frames (circular shape-retaining stiffeners) and stringers (length-wise stiffeners). The latter are equidistantly considered and fully fixed at their connections according to a fully integrated design. For the outer skin, the floor and the stiffeners, a shell formulation (9-node quads) combining a Reissner-Mindlin shell and a classical disc is considered. Normally, an optimal thickness distribution (outer skin and stiffeners) is generated in early aircraft design based on several load cases \cite{westphal2008}. For simplicity and due to a dimensioning not yet existing in the project, a constant thickness of $3\,$mm is assumed for all airframe structures, which lies in the range of a typical dimensioning for such aircraft sizes. In addition, for the frames, an I-shape with a constant height of $0.1\,$m is assumed and for the stringers, an U-shape with a constant height of $0.05\,$m is considered. In total, the introduced dimensions will definitely influence the occurring wavelengths and therefore the cabin \ac{SPL}, but are clearly not expected to influence the scientific findings with regard to solution efficiency significantly. Finally, structural damping is considered by a damping loss factor $\eta_1(f)$ based on measurements on \ac{CFRP} plates in \cite{dissBlech22}. In  $\Omega_1$, a complex-valued stiffness matrix (non-Hermitian) is therefore yielded.

The trim comprises the \textbf{insulation} $\mathbf{\Omega_2}$ and the \textbf{interior lining} $\mathbf{\Omega_3}$. For the insulation, the double wall gap is assumed to be completely filled by aircraft grade glass wool. An equivalent fluid approach (Helmholtz domain with complex material parameters) is chosen in combination with the Johnson-Champoux-Allard (JCA) model \cite{champoux1991dynamic,johnson1987theory} in order to derive the frequency-dependent required complex input parameter speed of sound $\underline{c}(f)$ and density $\underline{\rho}(f)$. This consequently leads to complex stiffness and mass matrices in $\Omega_2$. In addition, a limp frame extension is considered for the JCA density term \cite{panneton2007comments}. Based on experimental data, the JCA model is shown to be suitable with certain restrictions at low frequencies \cite{dissBlech22}. For the interior lining, honeycomb sandwiches combining a lightweight core and thin face sheets made of glass fibre reinforced plastics (GFRP) are commonly used \cite{ebnesajjad2010handbook}. For the model, a homogenised 3D continuum (27-node hexahedrons) and the above introduced 2D shell formulation (9-node quads) are assumed for the core and the face sheets, respectively.  Again, based on experimental data, material parameters for both the face sheets and the core are derived in \cite{dissBlech22} and applied to the underlying studies. Similarly to the airframe, structural damping is considered as well. 

Completing the model, the \textbf{cabin domain} $\mathbf{\Omega_4}$ is strongly coupled to the interior linings and the floor and finally delivers the sound pressure field in the cabin. A Helmholtz domain (27-node hexahedrons) is considered to model acoustic waves in the cabin. Damping by passengers and seats are introduced by a damping loss factor $\eta_4(f)$, the determination of which has been conducted by measurements in a reverberation chamber \cite{dissBlech22}. Again, the system matrix gets complex by introducing $\eta_4(f)$ in $\Omega_4$. \\

After assembling the FE element matrices, a linear FE system as given in Eqn.~(\ref{eq:systemDis}) comprising a global stiffness matrix $\mathbf{K}$ and mass matrix $\mathbf{M}$ is yielded.
\begin{equation}
	\underbrace{\left[\mathbf{K}-\omega^2\mathbf{M}\right]}_\mathbf{A}\mathbf{x}=\mathbf{f} 
	\label{eq:systemDis}
\end{equation}

The aircraft model setup delivers a certain structure within the matrices. Applying a consecutive renumbering to the domains $\Omega_{1-4}$, the submatrices depicted in Eqn.~(\ref{eq:detailMat}) and (\ref{eq:detailVec}) are resulting. 
\begin{alignat}{4}
	\mathbf{K} &= 
	\begin{bmatrix}
		\mathbf{K}_1 & -\mathbf{C}_\mathrm{12} & \mathbf{0}                       & -\mathbf{C}_\mathrm{14} \\
		\mathbf{0}            & \mathbf{K}_2            & -\mathbf{C}_\mathrm{23} & \mathbf{0}                       \\
		\mathbf{0}            & \mathbf{0}                       & \mathbf{K}_3            & -\mathbf{C}_\mathrm{34} \\
		\mathbf{0}            & \mathbf{0}                       & \mathbf{0}                       & \mathbf{K}_4            \\
	\end{bmatrix} &&,~
	\mathbf{M}&&=
	\omega^2
	\begin{bmatrix}
		\mathbf{M}_1                                     & \mathbf{0}                                                & \mathbf{0}                                                & \mathbf{0} \\
		\rho_\mathrm{2}\mathbf{C}_\mathrm{12}^\mathrm{T} & \mathbf{M}_2                                     & \mathbf{0}                                                & \mathbf{0} \\
		\mathbf{0}                                                & \rho_\mathrm{2}\mathbf{C}_\mathrm{23}^\mathrm{T} & \mathbf{M}_3                                     & \mathbf{0} \\
		\rho_\mathrm{4}\mathbf{C}_\mathrm{14}^\mathrm{T} & \mathbf{0}                                                & \rho_\mathrm{4}\mathbf{C}_\mathrm{34}^\mathrm{T} & \mathbf{M}_4 \\
	\end{bmatrix} \label{eq:detailMat} \\
	\mathbf{x} &=
	\begin{bmatrix}
		\mathbf{u}_\mathrm{1} \\
		\mathbf{p}_\mathrm{2} \\
		\mathbf{u}_\mathrm{3} \\
		\mathbf{p}_\mathrm{4} \\
	\end{bmatrix}
	\begin{tikzpicture}[baseline={([yshift=-9pt]current bounding box.center)},vertex/.style={anchor=base}]
		\draw[dotted] (0,-0.3) -- (0.3,-0.3) -- (0.3,0.6);
		\draw[dotted] (0.,0.6) -- (0.3,0.6);
		\node[inner sep=0pt, align=left, anchor=west] (hint) at (0.35,0.25) {\tiny shared \\[-9pt] \tiny nodes};
	\end{tikzpicture} &&,~
	\mathbf{f} &&=
	\begin{bmatrix}
		\mathbf{f}_\mathrm{ext} \\
		\mathbf{0} \\
		\mathbf{0} \\
		\mathbf{0} \\
	\end{bmatrix} \label{eq:detailVec}
\end{alignat}

\begin{figure}[!htb]
	\centering
	\begin{subfigure}[h]{0.29\textwidth}
		\centering
		\includegraphics[height=5cm]{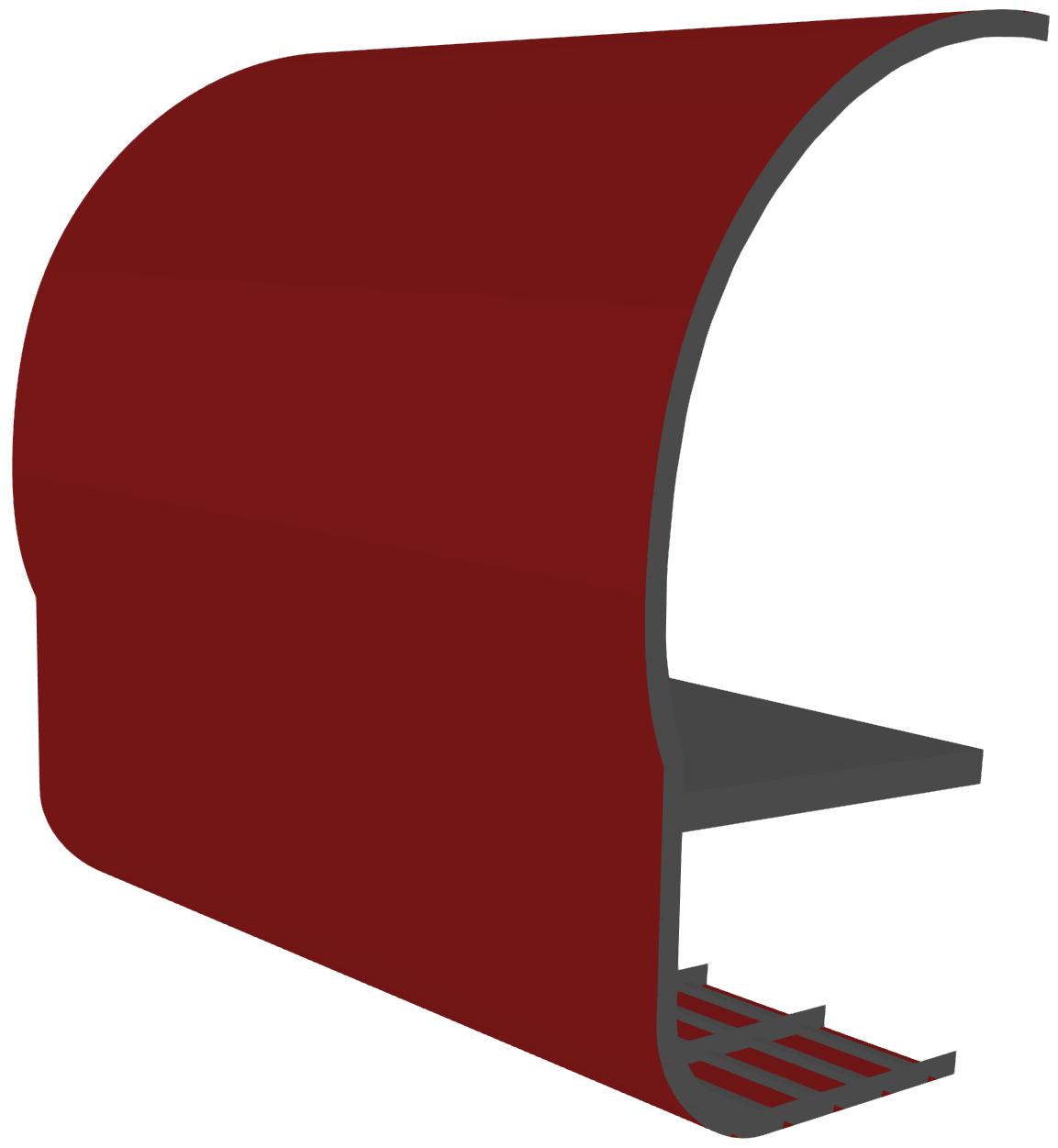}
		\caption{10 Hz}
	\end{subfigure} 
	\begin{subfigure}[h]{0.29\textwidth}
		\centering
		\includegraphics[height=5cm]{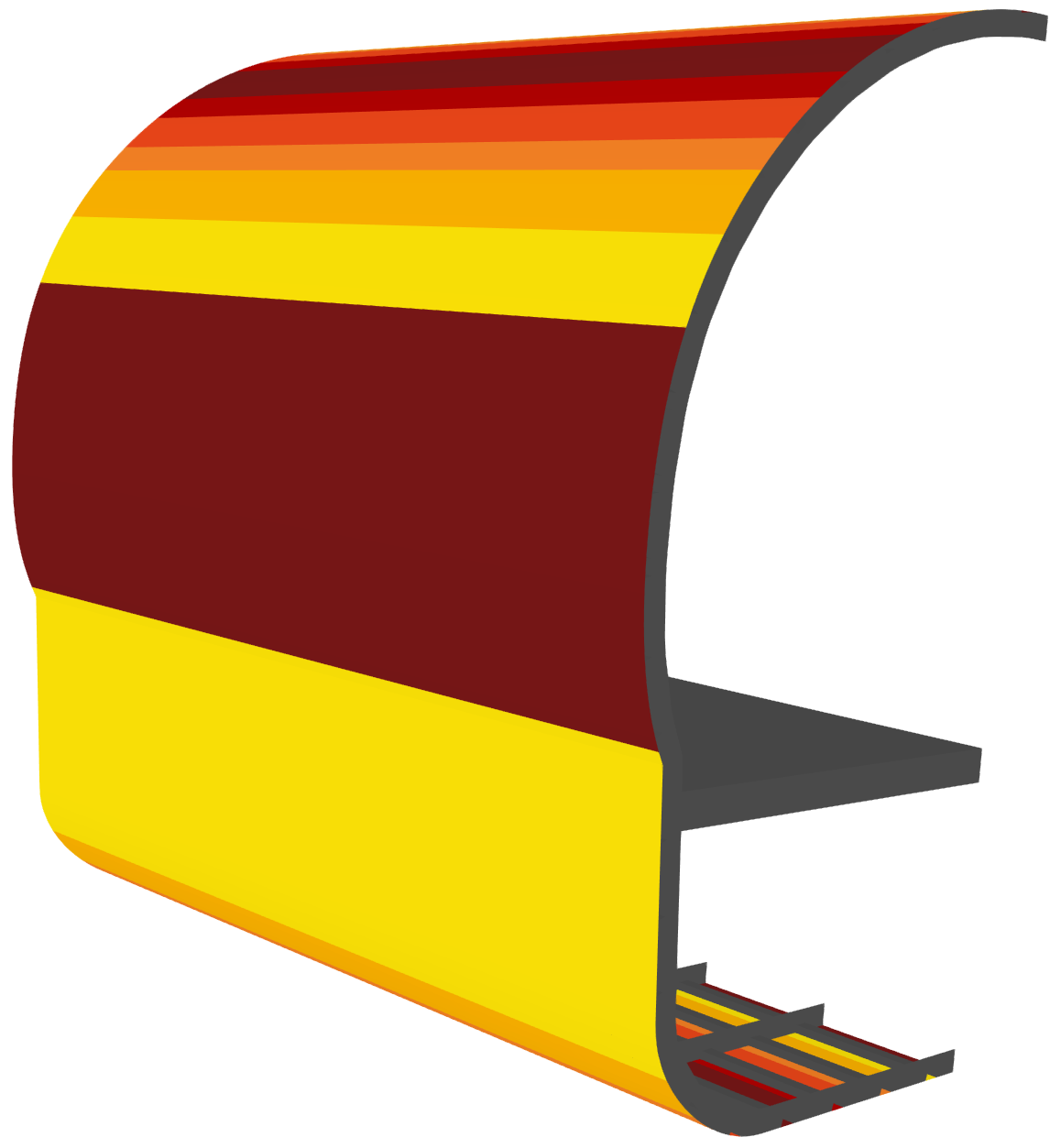}
		\caption{500 Hz}
	\end{subfigure} 
	\begin{subfigure}[h]{0.29\textwidth}
		\centering
		\includegraphics[height=5cm]{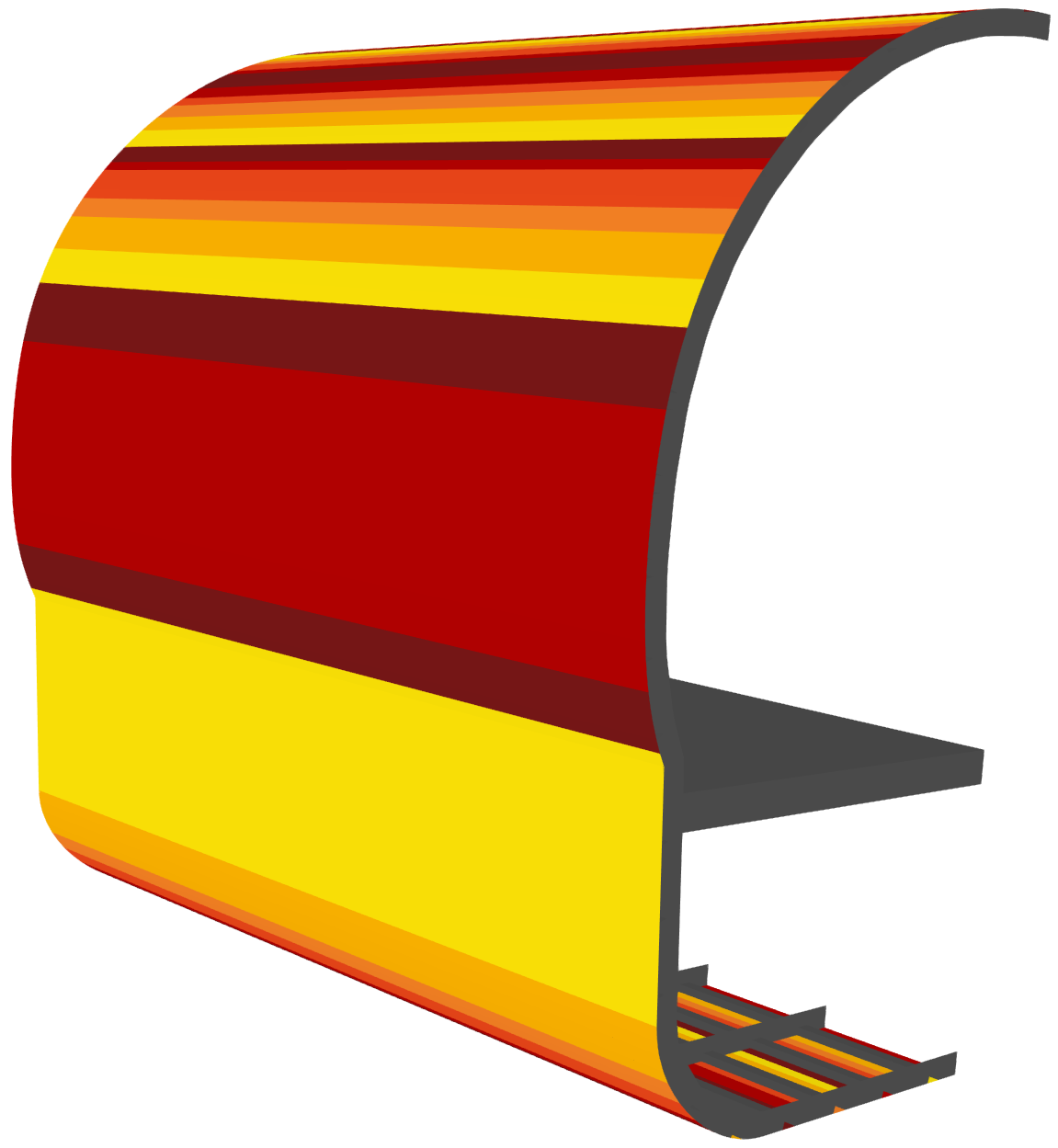}
		\caption{1000 Hz}
	\end{subfigure}
	\begin{subfigure}[h]{0.1\textwidth}
		\begin{tikzpicture}
			\begin{axis}[hide axis,colorbar,colormap name=warm,height=5.8cm,
				colorbar style={yshift=-2cm,
					yticklabels={$0$,$\pi$,$2\pi$},
					ytick={0,3.14,6.28},
					rotate=0,
					yticklabel pos=right,title=Phase (rad)},
				point meta min=0,
				point meta max=6.28,
			]
			{
			};
		\end{axis}
	\end{tikzpicture}
	\caption*{}
\end{subfigure}
\caption{Phase distribution of the frequency-dependent pressure wave load acting on the outer wall of the fuselage airframe domain $\Omega_1$ plotted for 10, 500 and 1000 Hz respectively}
\label{fig:freq_dep_load}
\end{figure}

For each domain, separate stiffness and mass matrices exist on the main diagonal of the system matrix. A fixed connection between the airframe $\Omega_1$ and the interior lining $\Omega_3$ (fixed joint at edges) is considered, while the two acoustic domains $\Omega_{2/4}$ are connected by coupling conditions introduced by coupling matrices $\mathbf{C}$. This way, displacements $\mathbf{u}$ induce sound pressures $\mathbf{p}$ and vice versa. By the consideration of damping loss factors for both structural and acoustic domains and coupling matrices, non-symmetric complex matrices and thus non-Hermitian system properties must be solved. For the studied aircraft application, a dominating frequency-dependent pressure excitation due to the turbulent boundary layer and propeller noise is expected on the outer skin as in Fig.~\ref{fig:freq_dep_load}. Therefore, the right hand side $\mathbf{f}$ in Eqn.~(\ref{eq:systemDis}) considers an excitation in $\Omega_1$ only, which is considered as a plane wave for this study. A replacement of the excitation is clearly expected without significant changes in the insights with regard to an efficient solution strategy and its transferability to different aircraft.

The frequency range of interest is limited at $1000\,$Hz for this work, which noticeably pushes the estimated practice oriented frequency limit of $200\,$Hz \cite{peiffer2016full}. An efficient sound transmission is expected within this range, though contributions by the turbulent boundary layer are relevant up to $3000\,$Hz. The latter is shown based on in-flight measurements in \cite{hu2013contributions}. Besides, a significant contribution of jet engines at several hundred Hertz as the second dominating source is depicted in \cite{hu2013contributions} as well. Nevertheless, the numerical prediction of turbulent boundary layers should be aimed for in future as jet engines are expected to become much quieter and therefore less dominating in future \cite{blech20springer}. However, based on the experiences of the authors, frequencies above $1000\,$Hz are currently hardly accessible for full aircraft, especially not for parameter studies.

When considering the vibroacoustic characteristics of an aircraft the surrounding conditions are of utmost importance. In order to assume stationarity, which helps simplify underlying mathematical equations, the model is to be examined in the cruise-configuration. It becomes obvious that an aircraft in cruise is subjected to pressurisation of the cabin, which leads to significant pressure differences between the inner cabin domain and the outer skin of the fuselage. As stated in \cite{dissBlech22}, the resulting overpressure affects the \ac{SPL} by introducing additional stiffness through the pre-stress of the aircraft's outer skin. Therefore, the pressure difference can be considered by applying a pre-stress to the fuselage elements \cite{dissBlech22}. With the assumption of pressurising a closed cylinder, the directional pre-stresses can be computed according to the following equation \cite{nilssonvibroacoustics}
\begin{equation}
\begin{gathered}
	T_{x}=\Delta p \frac{R}{2},\\
	T_{y}=\Delta p R.
\end{gathered}
\end{equation}

Hereby, $R$ is the radius of the closed cylinder. Considering the short range electrical propeller aircraft the overpressure $\Delta p$ is computed by assuming a cruising altitude of $7300\,$m. Subsequently, the overpressure yields $\Delta p = 42524\,$Pa. With a radius of $1.37\,$m, the resulting pre-stresses are $T_x= 27641\,$Pa and $T_y=55281\,$Pa, which we apply in all calculations. 

Finally, it becomes possible to solve the linear system of equations obtained from the FEM shown in Eqn.~(\ref{eq:detailMat}). In order to conduct efficiency studies on the presented aircraft model, the time spent as well as the memory used in the solving process are the most important indicators. Both these factors are also directly proportional to the system of equation's size, meaning the \ac{DoF}. Furthermore, the \ac{DoF}s depend on the mesh size or discretisation of the model. \\
The \ac{FEM} is a wave-resolving numerical method, meaning it can be used to solve for transversal and longitudinal waves propagating through the different domains of the aircraft. The main goal of utilising a wave-resolving numerical method is to depict the wave propagation accurately, also including the wave in itself. Therefore, a fitting discretisation has to be chosen, so that there are enough supports in the spatial domain for the wave to be depicted. Keeping the Shannon-Nyquist theorem in mind, it becomes obvious that the sampling frequency has to be at least double the highest frequency wave we want to depict in our computations. In the vibroacoustic problem presented here, especially in Eqn.~(\ref{eq:detailMat}), this sampling frequency can be directly related to the wavelength in different domains of the aircraft and therefore also to the spatial discretisation, since the problem is already presented in the frequency domain. In order to accurately depict a wave propagating through the present model, we choose to have at least 10 spatial supports per wavelength. This is the limiting factor for the discretisation, but in different domains the wavelength depends on specific material parameters such as speed of sound and density. Therefore, to be able to depict all waves accurately, the smallest wavelength present in the model and frequency domain serve as meshing criteria for the reference model. All domains are meshed by the same mesh size and the obtained model is considered for the entire frequency domain. It is possible to consider domain-adaptive meshing, which is the subject of Section \ref{sec:AdaptiveDiscretisation}. In order to show the full extent of the efficiency increases a conformingly meshed model is evaluated first. For the model depicted in Fig.~\ref{fig:modelOverview}, this gives us 2,434,743 \ac{DoF}s. The goal is to gain information on the solving process of this system and define this as a reference configuration, from which we can start utilising efficiency strategies to lessen computational effort. 

The direct solution conducted by the multifrontal direct solver MUMPS \cite{MUMPS:1, MUMPS:2} is set as reference. \cite{Sreekumar2021} provides a comparison of various parallel direct solvers solving linear systems from large-scale vibroacoustic problems in high-performance computing applications and \cite{Huepel2023} provide further insights into the effect of preconditioning with respect to various reordering schemes. All timings are conducted on the same system (Intel(R) Xeon(R) Gold 6138 CPU @ 2.00GHz, 40 physical cores, 768 GB RAM) using the vibroacoustic \ac{FEM} implementation elPaSo \cite{elpaso2023}. Several tests on the parallelisation setting concluded that 4 message passing interface (MPI) processes with 10 openmp (OMP) threads each lead to the fastest solving times of the system of equations and best memory utilisation with the direct LU solver on the used system. For the final reference, the discretisation in frequency domain is also of importance since the underlying system has to be solved for every frequency step. In a range from $10\,$-$1000\,$Hz, a $\Delta f$ of $2\,$Hz is chosen based on convergence tests, which results in $496$ frequency steps. The frequency step size is chosen optimally so as to sufficiently capture the occurrence of important resonances in the final \ac{FRF}. Between the different frequency steps, the performance of the direct solver does not change significantly, meaning that in order to compute the total time needed for the solve, $10$ frequency steps are solved and the mean time for solve and factorisation is taken. This mean time is then multiplied by the total number of frequency steps to obtain the total computation time.

For the configuration and size presented here, the mean solving time is $300.3\,$s per frequency step, leading to an overall computation time 
$42.1\,$h. This time is taken as the reference solution time where efficiency strategies can be employed, in order to lower the computational effort. 

\section{EFFICIENT SOLVING STRATEGIES}
\label{sec:strategies}

The goal and main advantage of precise aircraft simulations is the information on the \ac{SPL} level in an early design phase, before any prototype has been manufactured and tested. Engineers can have insights into cabin noise and derive design suggestions based on the cabin noise predictions. With simulations, many different material and parametric combinations can be tested more easily than with real environment testing. However, large-scale models, as also present in this paper, seem to reduce this rapid data evaluation by entailing significant computational effort. The introduced model of a fully electric regional aircraft might still be subjected to minor changes throughout the design phase and therefore also entails uncertainties. In order to evaluate many different configurations, lots of simulations have to be conducted. Nonetheless, a simulation time of $42$ h for just one of these configurations is neither adequate nor feasible. Therefore, the efficiency of computations has to be increased, so that the main advantages of simulations can still be upheld and many different aircraft designs can be simulated.

There exist a variety of methods to increase the efficiency of computations and decrease the entailing computational effort. Within this paper the focus is laid on three major methods, that the authors have broad experience with. 
Firstly, the size of the system of equations is decreased by utilising characteristics of the \ac{FEM} and applying domain-adaptive and frequency-dependent discretisations, focused on in \textbf{Sec.~\ref{sec:AdaptiveDiscretisation}}. Based on the optimally chosen meshes, two further improvements are paths are followed.

First, the solving process itself is adapted to the non-conforming meshes in \textbf{Sec.~\ref{sec:DD}}. By using the block properties of the underlying system of equations, domain decomposition approaches are investigated for an increase in solution efficiency in terms of solving time and memory requirements.

Second, \ac{MOR} techniques are applied in \textbf{Sec.~\ref{sec:mor}}. Recently, different mathematical methods, such as \ac{MOR}, have steadily made their way into structural mechanics as well as acoustics to enable faster computations. We reduce the size of the fuselage system by projecting the matrices on a low-dimensional mathematical subspace also termed as the projection-based \ac{MOR} techniques. Therefore, the resulting equations entail less computational effort. Recent literature dealing with \ac{MOR} methods in general as well as in vibroacoustics have always applied these methods to problems significantly less complicated than the aircraft model presented here. The novelty of successfully applying a \ac{MOR} method to such a large-scale model with coupled heterogeneous domains, while not decreasing the accuracy is shown in this section.

Finally, this section altogether gives an overview of the application of the above mentioned methods to the presented aircraft model and the results of decreased computational effort are shown. The proposed two ways are presented as potential workflow in order to cut down the overall computation time, so that it becomes feasible to compute the \ac{SPL} for many different aircraft configurations.

\subsection{Domain-adaptive and Frequency-dependent Discretisation}
\label{sec:AdaptiveDiscretisation}
When examining \textbf{domain-adaptive discretisation}, there are two main approaches. The first one was developed in \cite{Babuska.1975} and deals with adaptive mesh discretisation for specific subdomains. Here, the mesh can be refined in areas where a fine discretisation becomes necessary, e.g. when large gradients of the solution occur, while the rest of the mesh is kept fairly coarse. However, this contribution focuses on another domain-adaptive discretisation approach. 

Contrary to the \ac{DD} methods introduced in the following section, which aim to efficiently substructure and solve a given system of equations \cite{Smith.2015}, the methods presented in this section can be regarded as different \ac{DD} methods \cite{Maday.1988}. In these techniques, the domain is also split up in several different subdomains depending for example on physicality. However, since the substructuring is done before any discretisation process and the subdomains are treated independently, the underlying system of equations becomes smaller, meaning that the following section describes methods for efficiently solving systems of a set size, while the methods presented here aim to reduce the system size altogether. The approaches can be summarised as non-conforming discretisations or non-conforming meshes.\\
The main idea of non-conforming meshes is to allow different domains to be meshed independently from one another and the assembly of system matrices through special methods allows for an accurate coupling of the differently meshed domains. Since the special methods act like a mortar between the subdomains, these approaches are typically regarded as mortaring \ac{FEM}. In order to compute the \ac{SPL} arising from an external excitation in aircraft fuselages, a vibroacoustic model has to be used, meaning that a structure and an acoustic fluid have to be coupled as introduced in Sec.~\ref{sec:problem}. The dimension of the fluid is significantly larger than the structure's and therefore the main influence on the \ac{DoF}s stems from the discretisation of the fluid. If both subdomains are meshed conformingly it can lead to prohibitive system sizes and therefore, the non-conforming approach described previously can help reduce the computational effort by allowing for the fluid to be meshed coarser than the structure, leading to significantly less \ac{DoF}s. Additionally, the physicality of the vibroacoustic coupling allows for a simplified implementation of the non-conforming approach, meaning that with regards to the conforming implementation only the coupling matrices have to be adapted. The coupling matrix in a non-conforming case can be computed according to \cite{Kaltenbacher.2015}
\begin{equation}
	\mathbf{C}_{p,q}=\int_{\Gamma_{I}}^{}\rho_{f}\mathbf{N}^{s}_{p}\mathbf{N}^{f}_{q}\mathbf{n}d\mathbf{\Gamma},
	\label{CouplingMatrix}
\end{equation}
where $\Gamma_{I}$ describes the interface between the two subdomains, $\rho_{f}$ is the density of the acoustic fluid, and $\mathbf{N}^{s}$, $\mathbf{N}^{f}$ depict the ansatzfunctions of the structure and fluid respectively. The normal of the interface is denoted by $\mathbf{n}$. When comparing Eqn.~\eqref{CouplingMatrix} to the conforming implementation, there is almost no difference. However, the main challenge in implementing the previous equation is the numerical evaluation of the integral. For the computation of the integral, the Gaussian quadrature method is used. When evaluating the conforming integral in the parent space, the common Gauss points can be used, since the interface elements on both subdomains have the same local coordinates. This changes when evaluating the non-conforming integral. Since the elements of fluid and structure can overlap in any way, the identification of the Gauss points becomes the main challenge of implementing the non-conforming approach. Therefore, so-called interface elements are introduced and their local Gauss points are transformed into global coordinates. In order to transform the global Gauss points into the respective local coordinates of structure and fluid, a Newton algorithm becomes necessary \cite{SimonTriebenbacher.2010}. Finally, the entries in the coupling matrix can be computed according to \cite{SimonTriebenbacher.2010}
\begin{equation}
	\int_{\Gamma_{I}}^{}\rho_{f}\mathbf{N}^{s}_{p}\mathbf{N}^{f}_{q}\mathbf{n}d\mathbf{\Gamma}=\rho_{f}\sum_{e=1}^{n_{insec}}\sum_{l=1}^{n_{GP}}W_{l}\mathbf{N}^{s}_{p}(\xi^{s}_{l})\mathbf{N}^{f}_{q}(\xi^{f}_{l})\mathbf{J}^e(\xi^{e}_{l}).
	\label{EvalIntegral}
\end{equation}
In Eqn.~\eqref{EvalIntegral} $ {n_{insec}}$ and ${n_{GP}}$ denote the number of non-conforming interface elements and the number of Gauss points respectively. $W_{l}$ is the weight belonging to the respective Gauss points, while $\xi^{j}_{l}$, $j\in \{s,f,e\}$ are the local coordinates of the Gauss points in the respective parent space of the structure, fluid, and interface element. The Jacobi determinant is denoted by $\mathbf{J}$. An adequate implementation focuses on the identification of interfaces and the computation of accurate Gauss points for the evaluation of Eqn.~\eqref{EvalIntegral}. 

\begin{figure}[htb]
	\centering
	\input{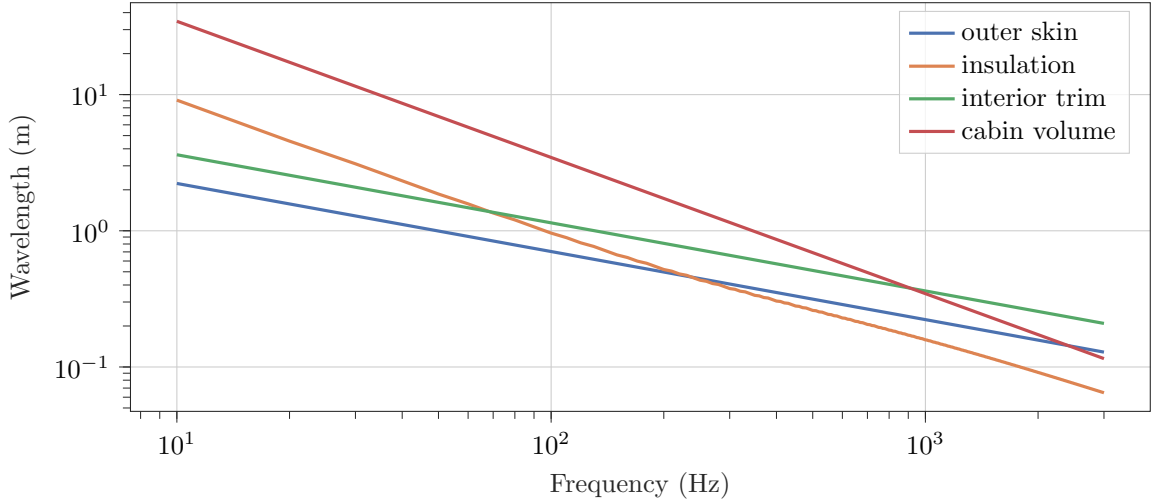}
	\caption{Depiction of the frequency-dependency of wavelengths in the four different domains}
	\label{fig:wavelengths}
\end{figure}

However, meshing different domains non-conformingly is not the only method to decrease the number of \ac{DoF}s. As already stated, discretisation depends on the wavelength with which the sound or structural bending waves are propagating through the system. But not only material parameters and different domains influence the wavelength. The wavelength is also highly \textbf{frequency-dependent}, meaning that the larger the examined frequency, the smaller the wavelength. As already mentioned above, when choosing an accurate discretisation, the smallest wavelength plays an important role, since it is governed by the largest frequency in the examined frequency domain and therefore directly influences the \ac{DoF}s. A general rule states the higher the examined frequency, the smaller the discretisation, the larger the system of equations. Therefore, frequency domains with large upper frequencies need large models to accurately depict the sound and structural bending waves. When examining a frequency domain from $10$-$1000$ Hz, as it is the case in this contribution, large models are entailed just by the minimum wavelength present. For better oversight of how sound and structural waves are propagating, the wavelengths over the frequency for the four main domains of the aircraft are plotted in Fig.~\ref{fig:wavelengths}. 

\begin{figure}[htb]
	\centering
	{%
		\setlength{\fboxsep}{0pt}%
		\setlength{\fboxrule}{0pt}%
		\fbox{\input{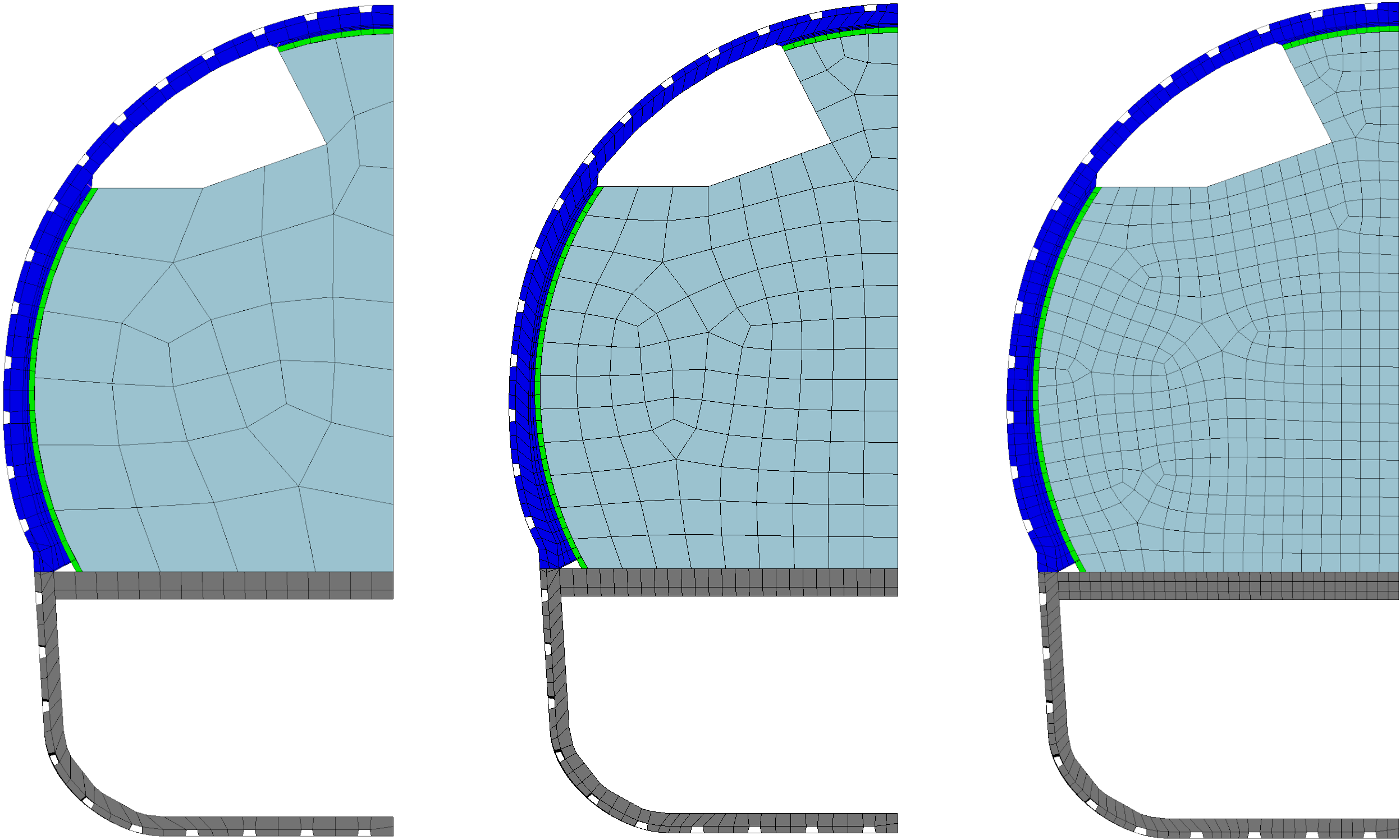}}}	
	\caption{Frequency- and domain-adaptive fuselage meshes. From left to right: Model used up to 258~Hz, 578~Hz, and 1000~Hz respectively. The mesh become finer in every domain with increasing frequency. The red mark on the left mesh indicates the discretisation criterion of elements per stringer distance. }
	\label{fig:FEmeshes}
\end{figure}

When examining Fig.~\ref{fig:wavelengths}, it becomes obvious that the discretisation is governed by the smallest wavelengths present, which is the outer skin of the aircraft for a low frequency range, where later on the smallest wave can be found in the insulation, at which point this becomes the main influence for choosing adequate mesh sizes. Furthermore, it becomes obvious that the figure showcases wavelengths up to $3000$ Hz. This serves to indicate the general behaviour of the waves in the different domains. The slopes are very different and while the structural parts govern the discretisation in lower frequency ranges, the slope of the cabin volume is the steepest, meaning that for higher ranges, finally, the cabin volume will be the main influence on discretisation. Generally, it becomes obvious why non-conforming meshes lead to a smaller system of equations, since the wave-resolving nature of the meshes can be obtained by meshing the subdomains differently, since there are different mesh sizes for every domain. However, the advantage of non-conforming meshes is expected to vary with increasing frequency. As a tendency, a decreased advantage can be expected, especially because of the behaviour of the sound waves in the cabin volume.\\
Nonetheless, it might not be feasible to only stick with one discretisation throughout the whole examined frequency domain. The mesh needed for $200$ Hz can be coarser than the mesh needed for $1000$ Hz. Therefore, a frequency-dependency can be introduced with the so-called frequency-dependent discretisation. Here, a criterion is chosen, so that a minimum requirement for discretisation is fulfilled. As soon as this criterion is broken a finer discretisation is chosen. In the case of this contribution, the frequency-dependent criterion is chosen as the minimum amount of nodes per wavelength. This also needs a base discretisation, for which the number of elements between stringers in the fuselage is chosen. Meaning that, as long as still feasible, the least amount of elements between stringers is chosen until they cannot longer guarantee ten nodes per wavelength, where another element in between the stringers is introduced, thus a finer discretisation is chosen. This leads to a frequency-dependent mesh, ultimately resulting in three different discretisations, which are shown in Fig.~\ref{fig:FEmeshes}.

The first mesh can be used from $10$-$258$ Hz, while the second mesh is used from $259$-$578$ Hz. Finally, the finest mesh is used from $579$ Hz up until the final $1000$ Hz chosen as the upper frequency in this contribution. This allows to cut \ac{DoF}s for more than half of the examined frequency domain. 

The number of \ac{DoF}s for the different domains and in combination for the model introduced in Sec.~\ref{sec:problem} are shown in Tab.~\ref{tab:freqdepMeshes}. It is important, that the finest discretisation is also the one used in the conformingly meshed reference configuration, since the same algorithms were employed to choose an adequate discretisation. Therefore, the frequency-dependent meshing helps in the lower frequency range. 

\begin{table}[htb]
	\caption{\ac{DoF}s, mean solution times $T$, memory requirements $M_\mathrm{LU}$ applying the reference MUMPS solver, and corresponding maximum frequency $f_\mathrm{max}$ for frequency-dependent meshing }\label{tab:freqdepMeshes} 
	\begin{tabular}{L{0.2\textwidth-12pt}C{0.2\textwidth-12pt}C{0.2\textwidth-12pt}C{0.2\textwidth-12pt}C{0.2\textwidth-12pt}}
		\toprule
		Elements between stringers & \ac{DoF} (1e6)\linebreak($\Omega_{1-3}$, $\Omega_4$) & $T\,$(s) & $M_\mathrm{LU}\,$(GB) & $f_\mathrm{max}$ (Hz)\\
		\midrule
		2 & 0.45 \linebreak (0.34, 0.12) & 13.5 & 11.3 & 258 \\
		3 & 1.05 \linebreak (0.60, 0.45) & 59.7 & 37.6 & 578 \\
		4 & 2.43 \linebreak (0.93, 1.5) & 300.3 & 164.3 & 1000 \\
		\bottomrule
	\end{tabular}
\end{table}

Furthermore, the time needed for solving each respective frequency step is shown in Tab.~\ref{tab:freqdepMeshes}. In order to compute the full solving time, the frequency steps in each range is multiplied by the average solving time given in the table. For the solely frequency-dependent discretisation the total solution time comes out to $21.6$ h, which cuts computation time down by more than half.

Insights into the total number of \ac{DoF}s are given in Tab.~\ref{tab:freqdepMeshes} as well. It becomes evident, that for the same accuracy in a lower frequency range, up to 2 million \ac{DoF}s can be saved, since they add no additional information to the solution process. Since this influences the size of the system of equations it should also affect the solution time. This can be observed in the table. However, computation time is not the only resource concerning efficiency. Another good indicator of efficient solving strategies is the memory used for the solution of the underlying system of equations. It is evident in Tab.~\ref{tab:freqdepMeshes}, that there is also a significant reduction of the memory requirements, further strengthening the usage of frequency-dependent meshing. As an indicator, we document the memory demand for the LU factorisation as the most expensive step and for a better comparability.

Still, even more efficiency gains can be reached by not only introducing frequency-dependent, but also domain-adaptive meshing, as explained above. This allows the cabin volume to be discretised coarser than the other domains and can save even more \ac{DoF}s and therefore lower the computational effort. The overview of \ac{DoF}s and average solving time for the frequency and domain-adaptive discretisations is shown in Tab.~\ref{tab:finalMeshes}. 

\begin{table}[htb]
	\caption{\ac{DoF}s, mean solution times $T$, memory requirements $M_\mathrm{LU}$ applying the reference MUMPS solver, and corresponding maximum frequency $f_\mathrm{max}$ for frequency-dependent meshing as well as domain-adaptive meshing.} \label{tab:finalMeshes}
	\begin{tabular}{L{0.2\textwidth-12pt}C{0.2\textwidth-12pt}C{0.2\textwidth-12pt}C{0.2\textwidth-12pt}C{0.2\textwidth-12pt}}
		\toprule
		Elements between stringers & DoF (1e6)\linebreak($\Omega_{1-3}$, $\Omega_4$) & $T\,$(s) & $M_\mathrm{LU}\,$(GB) & $f_\mathrm{max}$ (Hz)\\
		\midrule
		2 & 0.34 \linebreak (0.34, 0.00) & 5.6 & 5.3  & 258 \\
		3 & 0.63 \linebreak (0.60, 0.03) & 11.5 & 11.9 & 578 \\
		4 & 1.08 \linebreak (0.93, 0.15) & 36.2 & 29.2 & 1000 \\
		\bottomrule
	\end{tabular}
\end{table}

It can be seen in Tab.~\ref{tab:finalMeshes} that the structural parts are still discretised as in Tab.~\ref{tab:freqdepMeshes}, however, the cabin volume's \ac{DoF}s are further decreased significantly. This allows for a \ac{DoF} reduction by more than $50$\% for the finest discretisation, since the cabin is discretised as a three-dimensional domain, therefore giving the most opportunity for saving \ac{DoF}s.\\

When comparing Tab.~\ref{tab:freqdepMeshes} to Tab.~\ref{tab:finalMeshes}, it also becomes obvious that even though the mesh for $578$ Hz in Tab.~\ref{tab:freqdepMeshes} has about the same amount of \ac{DoF}s as the mesh for $1000$ Hz in Tab.~\ref{tab:finalMeshes}, the solution times, as well as the memory requirements are vastly different. This can be explained through the different structures of the underlying equations. Since in Tab.~\ref{tab:finalMeshes}, the cabin volume is meshed independently, there are not a lot of coupling terms, which changes the band structure of the matrices and therefore allows for a more efficient solve with the LU decomposition. Therefore, the differences in solution time and memory requirements are to be expected. Lastly, the domain-adaptive meshing in combination with frequency-dependent discretisation shows an improvement concerning the memory usage, where the required memory can be cut from $164.3$ GB to $29.2$ GB. 
Again, in order to compute the total solving time, the average solution time per frequency range is multiplied with the number of frequency steps in that region. The overall time for the fully utilised frequency- and domain-adaptive discretisation comes down to about $2.8$ hours. This shows a significant rise in efficiency when compared to the reference solution of $42.1$ hours (about $95$\%). Therefore, it is not only recommended to introduce adaptive meshing techniques when evaluating large-scale aircraft models, it would be not feasible to compute the reference solution without any adaption of the discretisation.
The choice of appropriate meshes is seen as the first essential step in order to increase the computational efficiency to the price of modelling effort. In the following, \ac{DD} and \ac{MOR} techniques are separately applied to the introduced meshes. 

\subsection{Domain Decomposition Techniques}
\label{sec:DD}

\begin{figure}[htb]
	\centering
	{%
		\setlength{\fboxsep}{0pt}%
		\setlength{\fboxrule}{0pt}%
		\fbox{\input{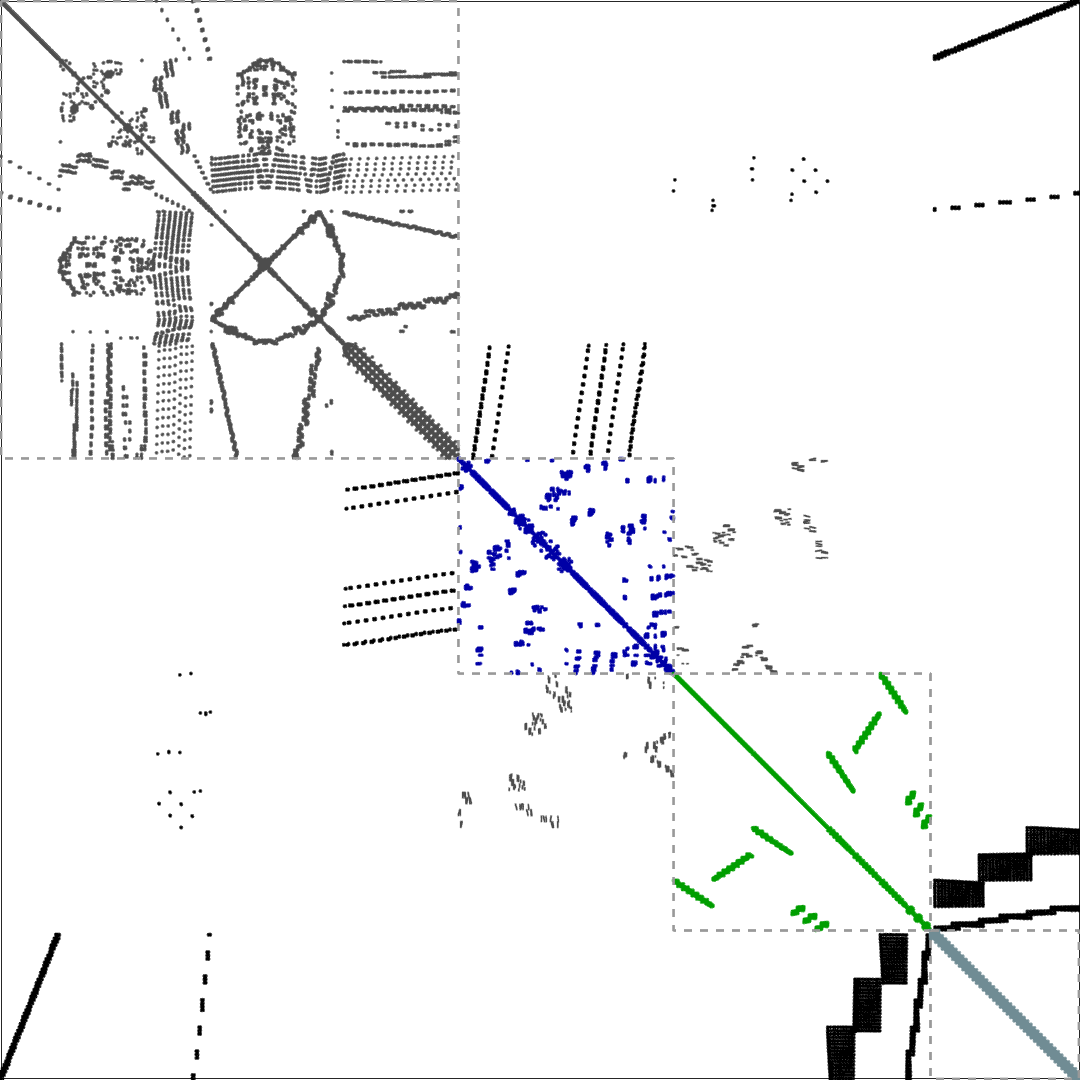}}}
	\caption{System matrix for the largest model (finest mesh in Fig.~\ref{fig:FEmeshes}) reordered according to physical domains}
	\label{fig:FEmatrices_sparsity1}
\end{figure}

As the discretised vibroacoustic system in Eqn.~(\ref{eq:detailMat}) has a natural block structure, \ac{DD} approaches are expected to be highly suitable. Each domain $\Omega_\mathrm{1-4}$ basically serves as one physical diagonal block, while the coupling entries are off-diagonal blocks \cite{poblet2010block}. Due to fundamentally different mechanical formulations, the structural and acoustic domains are numerically highly heterogeneous. Also due to this heterogeneous nature of the assembled system matrix, the individual physical domains introduce system entries with different order of magnitudes. As a result, the application of a suitable preconditioner has to be performed on each of the domains separately to alleviate the effort of any iterative solvers. For this reasons, we investigate performance gains by replacing the direct solvers with an iterative solver performing block or domain-wise preconditioning while using (a) non-overlapping \ac{DD} techniques like the block Jacobi preconditioner and (b) overlapping \ac{DD} techniques like the additive Schwarz method. As our different subdomains are strongly coupled, we expect an advantage for the iterative solvers to have access to the information from other domains by means of overlapping. As a result, the application of a suitable preconditioner has to be performed on each of the domains separately to alleviate the effort of any iterative solvers. Hence, we focus our further study towards physics-based \ac{DD} techniques that can handle solving our problems very well. In this context, physics-based \ac{DD} means the physical domains $\Omega_1$, $\Omega_2$, $\Omega_3$ and $\Omega_4$ are decomposed into respective subdomains. In Fig.~\ref{fig:FEmatrices_sparsity1}, the system matrix of the above-introduced FE mesh (finest model) is shown. The renumbering is chosen this way in order to apply \ac{DD} approaches for the different domains. As obvious, the airframe $\Omega_{1}$ is clearly dominating while the cabin domain $\Omega_4$ is relatively small due to the non-conforming meshing. In addition, coupling entries ($\mathbf{C}_{p,q}$) are clearly visible between most of the domains, which state the crucial cut-edges during decomposition and must be fit iteratively by the solver. 

Firstly, we investigate non-overlapping \ac{DD} techniques \cite{feng199933}, where an ideal monolithic preconditioner (LU factorisation) is applied to each domain. Such a block Jacobi preconditioner transfers the basic Jacobi preconditioner ($\mathbf{M}_\mathrm{pre}\coloneqq\mathrm{diag}(\mathbf{A})$) to blocks (domains) instead of diagonal entries \cite{saad2003iterative}. The block Jacobi preconditioner for a system of two non-overlapping blocks $\mathbf{A}_{1}$ and $\mathbf{A}_{2}$ is the sum of two individual preconditioners $\mathbf{M}_\mathrm{pre,1}$ and $\mathbf{M}_\mathrm{pre,2}$ as given in Eqn.~(\ref{eq:blockJac}) \cite{zhu2016generate}. Two independent (and therefore parallelisable) approximations for both $\mathbf{M}_\mathrm{pre,1}^{\mathrm{-}1}$ and $\mathbf{M}_\mathrm{pre,2}^{\mathrm{-}1}$ are possible. Again, the ideal preconditioner for each block $i$ is $\mathbf{A}_{i}^{\mathrm{-}1}$, e.g. obtained by a full LU factorisation. 

\begin{equation}
	\mathbf{M}_\mathrm{pre}=
	\begin{bmatrix}
		\mathbf{M}_\mathrm{pre,1} & 0 \\
		0 & \mathbf{M}_\mathrm{pre,2} 
	\end{bmatrix}
	\label{eq:blockJac}
\end{equation}

The combination of Eqn.~(\ref{eq:blockJac}) with the generalised minimal residual method (GMRES) is referred to as block Jacobi preconditioned Krylov subspace method \cite{zhu2016generate}. We expect \ac{DD} techniques to be competitive, especially for larger models. Hence, the investigation is focused on the most expensive model from Sec.~\ref{sec:AdaptiveDiscretisation}, which is the $1.08$e$6\,$\ac{DoF} model in Tab.~\ref{tab:finalMeshes}. 

In Tab.~\ref{tab:DDresultsLU}, we present some solver cases that yield converging solutions with good performance and are compared to the direct solver (MUMPS) as the reference configuration. For all the different solver cases, parallel distribution is kept fixed and corresponds to 4 MPI processes and 10 OMP threads per MPI process. An overview of the resulting computational times and memory demands is also shown in the table. For the memory consumption, the maximum demanded by the LU preconditioners is summarised according to the \ac{DD} configurations. Finally, to give an impression of the sufficient accuracy of the iterative solution updated in every iteration, we compute the relative error of the SPL in the cabin domain $\Omega$ and verify against the direct solver output where a set tolerance of $1$e$-2$ yields sufficient accuracy. 

\begin{table}[htb]
	\caption{Performance for various solver cases tabulating solution time per frequency step $T$, mean solution times $\bar{T}$ for 10 frequency step solves with solver object recycling and memory requirements $M_\mathrm{LU}$ for the $1.08$e$6\,$\ac{DoF} model ($f_\mathrm{max}=1000\,$Hz)} \label{tab:DDresultsLU}
	\resizebox{\textwidth}{!}{\begin{tabular}{lcccc}
		\toprule
		Solver case & $T$ (s) / Change  & $\bar{T}$ (s) / Change & $M_\mathrm{LU}$ $^1$ (GB) / Change & \multicolumn{1}{c}{\begin{tabular}[c]{@{}c@{}}Max. relative\\error $^2$\end{tabular}} \\ \midrule
		\multicolumn{5}{c}{Direct solver case as reference $^3$ - solving with 4 MPI and 10 OMP per MPI}                            \\ \midrule
		MUMPS            &  58 / - &  36 / -   &    29.2 / -     &                    - \\ \midrule
		\multicolumn{5}{c}{Configuration 1: Iterative solver - solving 4 subdomains [$\Omega_1,\Omega_2,\Omega_3,\Omega_4$] with 4 MPI and 10 OMP per MPI}                            \\ \midrule
		GMRES $^4$ with GASM $^5$ with 1 overlap & 42 / -28\% & 33 / -8\%    & 4.5 / -84\%          & $3.02\cdot 10^{-3}$                    \\
		GMRES $^4$ with GASM $^5$ with 2 overlaps & 64 / +10\% & 49 / +36\%     & 3.3 / -89\%             & $1.66\cdot 10^{-3}$                     \\
		GMRES $^4$ with block Jacobi $^6$ preconditioning & 137 / +226 \% & 123 / +272 \%   & 8.1 /  -72\%               &  $7.18\cdot 10^{-2}$                   \\ \midrule
		\multicolumn{5}{c}{Configuration 2: Iterative solver - solving 2 subdomains [$\Omega_1 \cup \Omega_2 \cup \Omega_3,\Omega_4$] with 4 MPI and 10 OMP per MPI}                            \\ \midrule
		GMRES $^4$ with GASM $^5$ with 1 overlap &  77 / +33\% &  57.8 / +60\%    &  10.1 / -65\%          & $3.44\cdot10^{-3}$                    \\  \midrule
		\multicolumn{5}{c}{Configuration 3: Iterative solver - solving 3 subdomains [$\Omega_1 \cup \Omega_3, \Omega_2,\Omega_4$] with 4 MPI and 10 OMP per MPI}                            \\ \midrule
		GMRES $^4$ with GASM $^5$ with 1 overlap &  60 / +3\% &   47.9 / +33\%    &  10.8 / -63\%          & $2.46\cdot10^{-3}$                    \\ 
		\bottomrule
	\end{tabular}} \\[0.1cm]
	$^1$ Mean memory consumption during the LU factorisation stage applicable to the whole matrix in case of direct solver and the decomposed subdomain matrix in case of iterative solver based on \ac{DD}  \\
	$^2$ Maximum relative error over 10 frequency steps for mean \ac{SPL} in cabin domain $\Omega_4$ with MUMPS direct solution as reference \\
	$^3$ Total time for solve for the first frequency step contain a one-time analysis phase \\
	$^4$ GMRES with \texttt{PETSc} settings: \\ \texttt{-ksp\_type gmres -ksp\_diagonal\_scale -ksp\_gmres\_modifiedgramschmidt -ksp\_gmres\_restart 1000 -ksp\_max\_it 150 -ksp\_atol 1e-4  -sub\_ksp\_type preonly -sub\_pc\_type lu -sub\_pc\_factor\_mat\_solver\_type mumps} \\
	$^5$ GASM with \texttt{PETSc} settings: \\
	\texttt{-pc\_type gasm -pc\_gasm\_type restrict -pc\_gasm\_total\_subdomains 4 -pc\_gasm\_overlap \#} \\
	$^6$ GASM with zero overlapping subdomains
\end{table}

Direct solvers, here the reference case in Tab.~\ref{tab:DDresultsLU}, perform LU decomposition by equally distributing the matrices and vectors to the 4 MPI processes. The process includes a symbolic factorisation stage, a factorisation phase where the LU decomposition is realised and a solution phase where forward and backward substitution is performed to find the system's solution. In the context of frequency domain analysis, the various frequency steps can be solved by reusing the symbolic factorisation for all subsequent solves and hence is performed only once. Therefore, the time for the first frequency solve including the time for symbolic factorisation is compared to the mean solving time averaged over the 10 frequency steps in Tab.~\ref{tab:DDresultsLU}. In PETSc, one can reuse the same solver object for multiple such frequency evaluations and avoid recomputing the symbolic factorisation stage. To represent the memory consumption of MUMPS, the maximum memory requirement in the factorisation stage is also recorded. Further details on the used direct solver settings and related investigations can be referred to \cite{Sreekumar2021,Huepel2023}.

Though direct solvers are efficient enough to solve the current problem as stated in Tab. \ref{tab:finalMeshes}, the search for feasible iterative solvers is underway to deal with very large-scale matrices and also provide an even faster solving technique. For the finest discretised fuselage system representing a heterogeneous system, we consider \ac{DD} techniques presented at the beginning of this section by combining GMRES with an additive Schwarz method. The additive Schwarz procedure is comparable to the block-Jacobi procedure when there is no overlapping of subdomains. The resulting preconditioner can be then expressed as $\mathbf{M}_\mathrm{pre,AS}=\sum_{i=1}^{n_b} \mathbf{R}_j^T \mathbf{A}_j^{-1} \mathbf{R}_j$ \cite{saad2003iterative} for $n_b$ number of subdomains and $\mathbf{R}_j$ are the individual residuals. As a block-preconditioner, a generalised implementation of the additive Schwarz method within PETSc (GASM) is considered. The GASM implementation allows for overlapping and subdomains arbitrarily spread over MPI ranks. Especially the latter is required in order to merge physical subdomains over several processes to fully exploit MUMPS' parallel capabilities. For each subdomain in all settings, a full LU factorisation is applied by the parallel MUMPS solver as an ideal preconditioner. This way, we mainly focus on the convergence rates resulting from the decomposition. As an iterative solver, GMRES is applied within PETSc to compute the exact solution with a set error tolerance. Especially for ill-conditioned vibroacoustic problems with heterogeneous domains, GMRES combined with physical domain-based preconditioning is a stable choice. For such problems, there are hardly any alternative solvers to consider due to instabilities or convergence problems \cite{dissBlech22}. And also like in the case of direct solvers, the GMRES solver objects in PETSc can be reused for multiple frequency evaluations for an improved starting guess and to accelerate the overall convergence.

A range of iterative solver settings performing \ac{DD} is possible for the considered aircraft model so as to yield converging results. Of the various configurations that we investigated, the best performing cases are presented in Tab.~\ref{tab:DDresultsLU} as configurations 1-3. Firstly, each of the physical domains $\Omega_{1-4}$ are distributed to each one of the 4 MPI processes each processed by 10 OMP threads yielding the configuration of 4 decomposed subdomains $[\Omega_1,\Omega_2,\Omega_3,\Omega_4]$. In high-performance computing clusters, this can be related to distributing individual domains to a number of computing nodes. A second configuration $[\Omega_1\cup \Omega_2\cup \Omega_3,\Omega_4]$ accounts for grouping of the strongly coupled airframe, thereby decomposing the system into two subdomains - each acted upon by 2 MPI processes. Finally, a third configuration $[\Omega_1 \cup \Omega_3, \Omega_2,\Omega_4]$ is formed by only grouping the closely coupled structural domains delivering a total of three subdomains where the two distinct acoustic domains $\Omega_2$ and $\Omega_4$ are treated separately. By comparing to the costs of a full MUMPS solution tabulated in Tab.~\ref{tab:finalMeshes}, we find a good convergence of the solution using GASM with 1 overlapping domain for the first configuration where all domains are treated separately. This setting provides a saving in time (almost 8\% reduction) and, most importantly, significant savings in terms of memory (84\% reduction) due to decomposed domains that are rather smaller when compared to the other two configurations and direct solving. Moreover, this splitting of the different physical domains ($\Omega_{1-4}$) is observed to significantly aid convergence. When the number of overlaps is increased, an increase in accuracy is observed at the cost of higher solving time and memory requirements. For the case with no overlap which corresponds to the classical block Jacobi preconditioning, the iterative solver is significantly slower, which implies the necessity of having more information from the coupled subdomains by means of subdomain overlapping.  On the other hand, configurations 2 and 3 yield an accurate solution when tested for the solver setting with 1 overlap but still require more effort when compared to the best configuration 1.

In the context of parallel computing, the optimal \ac{DD} presented above contains MPI processes working on various domains of different sizes. This evidently requires attention towards load balancing and is identified as an area for future performance optimisations. Also, we identify further extensions of the above study with respect to the number of subdomains and the number of MPI assigned to solve each of the subdomains. Also, recycling underlying Krylov subspaces \cite{JOLIVET2021277} is an interesting aspect that can be considered for solving systems at multiple frequency points. In addition, we have also conducted investigations to perform \ac{DD} based on the entire mesh and distribute numerical domains obtained by performing a reordering scheme on the entire system matrix. This approach has led to the least success due to preconditioning without physics-based domain treatment. So far, always full LU factorisations are applied as preconditioners in the individual domains. For a further acceleration of the computation time, cheaper preconditioners like incomplete LU are planned to be studied in future. Especially for the physics-based approach, which considers larger domains, incomplete factorisations might increase efficiency.

Summarising the section on \ac{DD} techniques, the mesh-based approach using GMRES using GASM preconditioners with overlapping subdomains shows the most promising results. Compared to the parallel MUMPS solver, the iterative setting with 4 subdomains and 1 overlapping improves efficiency by reducing the total computational time by a small margin of 8\% (from 2.8 hours in Sec.~\ref{sec:AdaptiveDiscretisation} setting to the current setting requiring 2.6 hours), but at a significantly lowered memory requirement of -84\%. As a result, \ac{DD} techniques are highly promising for solving even large-scale models.

\subsection{Rational Arnoldi Krylov Subspace Method}
\label{sec:mor}

\ac{MOR} techniques are a popular approach for an efficient approximation of large-scale systems \cite{Antoulas2005}. The method aims at the generation of an accurate low-order approximation of the expensive full-order system, with which computations can be executed faster in the reduced space without compromising on the accuracy of the sought solution.

In this section, state-of-the-art \ac{MOR} techniques are applied to efficiently reduce the system represented in Sec.~\ref{sec:problem}. We deploy the moment matching algorithm of the second-order \ac{rA-Krylov} \cite{SALIMBAHRAMI2006385,Bai2005,sreekumar2021efficient} to accurately approximate the system response within a broad frequency spectrum. Firstly, we present the vibroacoustic problem and the reduction framework to handle the presented large-scale models. Then we elaborate on the results obtained by reducing the aircraft models and finally we conclude this section with an intermediate conclusion by listing the actual challenges.

The dimensionality reduction for our vibroacoustic problems is performed on the second-order dynamic system of equations and the resulting full-order model can be expressed as:
\begin{eqnarray}
	\left( -\omega^2 \mathbf{M}(\omega) + i\omega \mathbf{D}(\omega) + \mathbf{K}(\omega) \right) \mathbf{x}(\omega) =\mathbf{f}(\omega) ;\, \mathbf{y}(\omega) = \mathbf{C}^T \mathbf{x}(\omega),
	\label{eq:FOM_FREQDEP}
\end{eqnarray}
where the system matrices $\mathbf{M}(\omega),\mathbf{D}(\omega),\mathbf{K}(\omega) \in \mathbb{C}^{n\times n}$ are frequency-dependent as a consequence of frequency-dependent material parameters, for instance Young's modulus. In addition, the input to the system $\mathbf{f}(\omega) \in \mathbb{C}^{n}$ is the frequency-dependent load which later represents the plane pressure wave excitation acting on the outer aircraft fuselage walls plotted in Fig.~\ref{fig:freq_dep_load}. $\mathbf{x}(\omega) \in \mathbb{C}^n$ denote the system state, $\mathbf{y}(\omega)\in\mathbb{C}^{n_o}$ the system output and $\mathbf{C}\in \mathbb{R}^{n\times n_o}$ the output matrix marking the output \ac{DoF} of interest. For the aircraft model, as mentioned in Sec.~\ref{sec:problem}, numerical damping is introduced as structural damping with frequency-dependent damping loss factor $\eta(\omega)$. As a result, comparing to the system of equations presented in Eqn.~\eqref{eq:systemDis}, the structural damping yield stiffness proportional damping matrix that varies inversely to the frequency such that $\mathbf{D}(\omega) = \eta(\omega) \hat{\mathbf{K}}/\omega$ where $\hat{\mathbf{K}}:=\mathbf{K}$ is the general stiffness matrix term. Or in other words, one can represent structural damping introduced as the complex-valued stiffness matrix, such that $\mathbf{K}(\omega) = \hat{\mathbf{K}}(1+i\eta(\omega))$, instead of using the damping matrix.

A conventional solving of the problem presented in Eqn.~\eqref{eq:FOM_FREQDEP} is computationally expensive and therefore a dimensional reduction with \ac{MOR} can yield faster results. However, the aircraft model is highly challenging for existing \ac{MOR} approaches due to the large number of inputs acting on the system and the frequency-dependent nature of its system matrices. Moreover, the considered vibroacoustic problem presents new challenges due to the high amount of dynamics or modal density owing to the fluid-structure interaction. As a result, we identify some special approaches to overcome these challenges to a certain extent, which are discussed below.

The second-order dynamic equation in Eqn.~\eqref{eq:FOM_FREQDEP} is reduced using the projection-based \ac{MOR} techniques, where the projection bases span the second-order Krylov subspace \cite{BAI2002, Bai2005} expressed as:
\begin{eqnarray}
	\mathrm{colspan}(\mathbf{V}) = \bigcup_j  \mathcal{K}(-\tilde{\mathbf{K}}^{-1}_j\tilde{\mathbf{D}}_j,-\tilde{\mathbf{K}}^{-1}_j\tilde{\mathbf{M}}_j, -\tilde{\mathbf{K}}^{-1}_j{\mathbf{f}}_j)
	\label{eq:SOKSM}
\end{eqnarray}
where $\tilde{\mathbf{K}}_j = -\omega_j^2 \mathbf{M}(\omega_j) + i\omega \mathbf{D}(\omega_j) + \mathbf{K}(\omega_i), \, \tilde{\mathbf{D}} = 2i\omega_j \mathbf{M}(\omega_j) + \mathbf{D}(\omega_j), \, \tilde{\mathbf{M}}=\mathbf{M}(\omega_j)$ and $\mathbf{f}_j=\mathbf{f}(\omega_j)$ are the shifted system matrices calculated at the various expansion points $\omega_j \in [\omega_{\text{min}}, \omega_{\text{max}}]$. The expansion points are chosen iteratively from the desired frequency region using the greedy approach. In case of the considered structural damping case where the damping is introduced as complex stiffness, the resulting system can be reduced with an equivalent first-order Krylov subspace where the damping matrix term in Eqn.~\eqref{eq:FOM_FREQDEP} vanishes or is not used \cite{sreekumar2021efficient}, such that:
\begin{eqnarray}
	\mathrm{colspan}(\mathbf{V}) = \bigcup_j  \mathcal{K}(-\tilde{\mathbf{K}}^{-1}_j\tilde{\mathbf{M}}_j, -\tilde{\mathbf{K}}^{-1}_j{\mathbf{f}}_j).
	\label{eq:FOKSM}
\end{eqnarray}

An efficient reduction for the considered vibroacoustic problem can be ensured with the classical moment-matching theorems \cite{EricJamesGrimme.,SALIMBAHRAMI2006385}, while accounting for the frequency-dependent nature of the system. A frequency-affine decomposition is not feasible here due to a large number of frequency-dependent material parameters in different domains $\Omega_{1-4}$. Using parametric \ac{MOR} \cite{benner2015survey} approaches for considering the system variation with respect to frequency would be relatively expensive and require approaches to handle the high-dimensional parameter space \cite{romeradaptive, Sreekumar2022}. Hence, classical \ac{MOR} approaches are still applied by computing moments corresponding to the respective full-order model matrices evaluated for the various expansion points. In this way, the frequency-dependent nature of the considered vibroacoustic problem in Eqn.~\eqref{eq:FOM_FREQDEP} can be essentially captured. Regardless, considering the frequency-dependent effects implicitly for moment computation can accelerate the \ac{MOR} convergence and is interesting for detailed investigations. Finally, for the presented aircraft model with frequency-dependent material parameters, the reduced system of equations is of the form:
\begin{eqnarray}
	\left( -\omega^2 \mathbf{M}_R(\omega) + i\omega \mathbf{D}_R(\omega) + \mathbf{K}_R(\omega) \right) \mathbf{x}_R(\omega) = \mathbf{f}_R(\omega) ;\, \mathbf{y}_R(\omega) = \mathbf{C}_R^T \mathbf{x}_R(\omega),
\end{eqnarray}
where $[\,\cdot\,]_R(\omega) = \mathbf{V}^H [\,\cdot\,]\mathbf{V}$ with $[\,\cdot\,]:= \left\{\mathbf{M}(\omega), \mathbf{D}(\omega), \mathbf{K}(\omega)\right\}$, $\mathbf{f}_R(\omega) = \mathbf{V}^T\mathbf{f}(\omega)$ and $\mathbf{C}_R = \mathbf{C}^T \mathbf{V}$. We term the resulting reduced system as the \ac{fROM}. Though the damping matrix is not used for our models as mentioned earlier due to the usage structural damping model, further damping can be still investigated in the form of Rayleigh damping with the existing fROMs using $\mathbf{D}_R = \alpha \mathbf{M}_R + \beta \mathbf{K}_R$ \cite{Eid2007}.

In the usual \ac{rA-Krylov} setting, a multiple-input-multiple-output configuration is required to account for exciting multiple \ac{DoF}s of the outer skin $\Omega_1$ in the aircraft fuselage. However, due to a large number of excited nodes and respective \ac{DoF}s, it is not feasible to consider such a large number of inputs and the resulting number of moment computations per greedy iteration that leads to tremendous memory requirements. As a trade-off between the computational complexity during this \ac{MOR} offline phase and the resulting ROM dimension, the dynamic loading term $\mathbf{f}(\omega)$ is considered explicitly for moment computation at the respective expansion points as expressed in Eqn.~\eqref{eq:SOKSM}. Though the yielded fROM is only applicable to the type of loading chosen, such an approach delivers efficient fROM computations in terms of the overall size of the fROM and faster MOR offline phase. Details on the Krylov subspace methods can be referred to \cite{EricJamesGrimme.} and \cite{sreekumar2021efficient} for the deployed \ac{rA-Krylov} algorithm.

Now that the background theory of \ac{MOR} been said, further on, we consider the aircraft model presented in Sec.~\ref{sec:problem} for a {reduction in system dimension}. The domain-adaptive and frequency-dependent discretisation adopted in Sec.~\ref{sec:AdaptiveDiscretisation} allows the usage of three models with different mesh sizes for various frequency regions ranging from 10 to 1000 Hz. Although these models are of relatively small dimensions when compared to the conventional model obtained from a conforming mesh implementation, they are still expensive. The dimension of the full system matrices and their respective solving time required per frequency solve, listed in Tab.~\ref{tab:finalMeshes}, motivates the requirement of a surrogate for faster computation in the frequency domain. Hence, we perform reduction to the three aircraft models with \ac{rA-Krylov} to obtain fROMs to circumvent the huge computational expense.

\begin{figure}[htb]
	\centering
	\begin{tikzpicture}
		\node (0,0) {\includegraphics[width=0.5\linewidth]{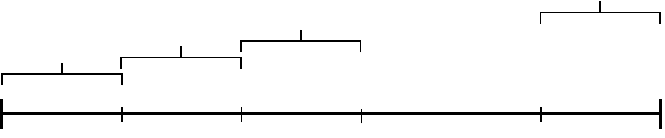}};
		\node at (-4.5,-1.1) {$\omega_{\text{min}}$};
		\node at (4.5,-1.1) {$\omega_{\text{max}}$};
		\node at (-3.6,0.5) {$\text{l-fROM 1}$};
		\node at (-2.,0.8) {$\text{l-fROM 2}$};
		\node at (-0.4,1.1) {$\text{l-fROM 3}$};
		\node at (3.6,1.5) {$\text{l-fROM n}$};
		\node at (1.4,1.25) {$\cdots$};
	\end{tikzpicture}
	\caption{Schematic representation of local fROM generation for broadband MOR simulations}
	\label{fig:l-from}
\end{figure}
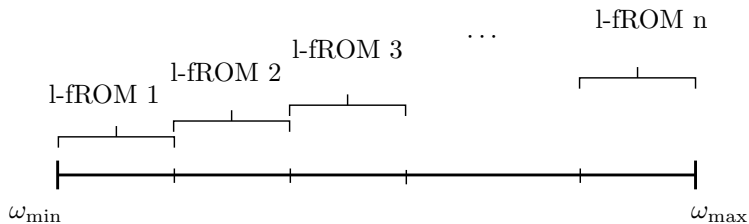

One major challenge when approximating fluid-structure coupled models is their increasing modal density at higher frequencies. As a consequence, the yielded solutions or \ac{FRF}s are highly dynamic in nature with a large number of resonances. Creating a global fROM valid for the respective frequency regions and for the considered aircraft model becomes challenging at higher frequency regions. A global fROM for a broader frequency domain also indicates larger fROM dimensions or requirement of a large number of moments which adds to the computational cost - both in \ac{MOR} offline and online phase. An alternative is to suitably partition the broad frequency domain into sub-intervals and generate local fROMs (l-fROM) for these frequency windows as illustrated in Fig.~\ref{fig:l-from}. Such an approach delivers relatively small fROMs and is able to approximate the highly varying FRFs as a result of the high modal density. Hence for higher frequencies, we generate l-fROMs for definite sub-intervals depending on the increasing modal density.

A summary of the three frequency-dependent meshes and the considered sub-intervals or frequency windows for (l-)fROM generation is presented in Tab.~\ref{tab:MOR_ModelOverview}. For model 1, valid at lower frequencies (10-258 Hz), a global fROM is created due to comparatively low modal density. But for model 2 (258-578 Hz) and model 3 (578-1000 Hz), we perform a partitioning of the frequency domain into a suitable number of windows where l-fROMs are generated. The results of performing \ac{MOR} with \ac{rA-Krylov} are also tabulated in Tab.~\ref{tab:MOR_ModelOverview}. 

\begin{figure}[htb]
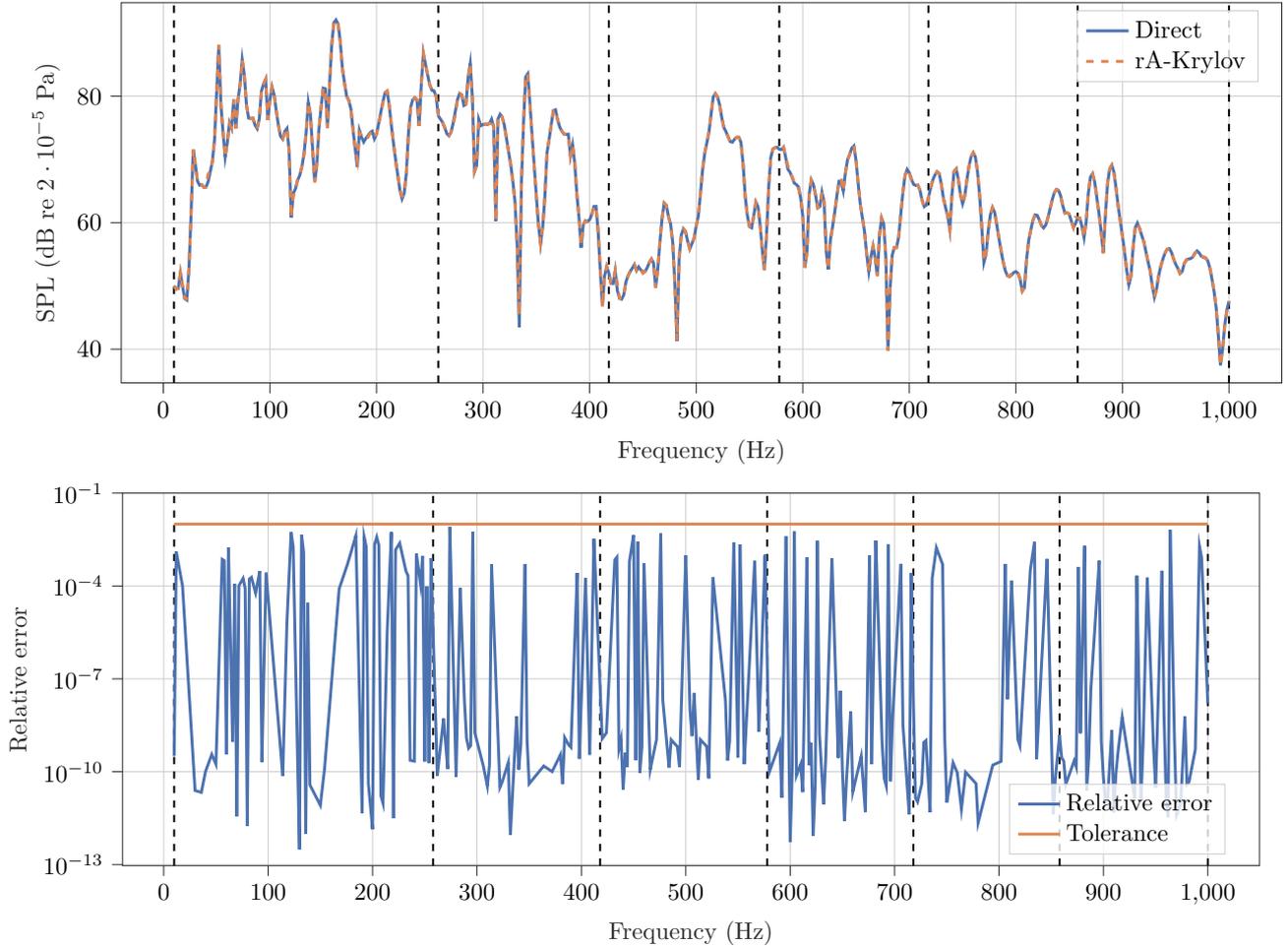

	\centering
	\begin{subfigure}[h]{\textwidth}
		\resizebox{\textwidth}{!}{\input{./images/pyplots/mor_tfs/plt_transferfunction_AIR_ALLDOM.tex}}
	\end{subfigure}
	\begin{subfigure}[h]{\textwidth}
		\hspace{-0.3cm}\resizebox{\textwidth}{!}{\input{./images/pyplots/mor_tfs/plt_overallerror_AIR_ALLDOM.tex}}
	\end{subfigure}
	\caption{Comparison of pressure \acp{FRF} (upper figure) for node at coordinates $(x,y,z)=(-0.428,1.228,1.500)$ consolidated from respective full order model and l-\ac{fROM} computations. The corresponding relative error for the same \ac{FRF} plot is also plotted (lower figure). Dashed lines denotes the frequency-interval boundaries for the generated l-\acp{fROM}.}
	\label{fig:mor_frf_error_consolidated}
\end{figure}

The major cost while performing \ac{MOR} incur to the \ac{fROM} generation during the offline phase. This accounts for the moment computation, according to Eqn.~\eqref{eq:FOKSM}, involving the full system matrix factorisation for multiple expansion points. Using MUMPS direct solver, the factorisation is stored for further moment computation at the same expansion point. Hence, the significant computational time for \ac{MOR} offline procedure is on par with performance values as per Tab.~\ref{tab:finalMeshes}. 
Moreover, due to a large number of eigenmodes present in the system with increasing frequency as tabulated in Tab.~\ref{tab:MOR_ModelOverview}, a high number of expansion points (around 20-40 expansion points) is required to sufficiently capture the dynamics. This adds to the overall computational expense in the MOR offline phase. In addition, we manually partitioned the frequency domain for obtaining optimal l-fROMs at higher frequencies. An automated method to perform an optimal partition depending on the nature and complexity of the vibroacoustic problem is interesting for future research. Regardless, we obtain accurate and converging fROMs of very small dimensions approximating the required system response with sufficient accuracy. See Fig.~\ref{fig:mor_frf_error_consolidated} for the comparison of pressure FRFs obtained at the desired node with the fuselage cabin consolidated with results obtained from various fROMs. As a result, a conventional solve incurring computational cost tabulated in Tab.~\ref{tab:finalMeshes} can be replaced with solving in the reduced space or termed as the \ac{MOR} online phase yielding significant speedup, as in Tab.~\ref{tab:MOR_ModelOverview}, due to the very small dimension of the generated fROMs. Such multiple frequency-domain surrogates delivering fast response computations can be further incorporated in the parametric MOR framework, refer to \cite{benner2015survey,romeradaptive, Sreekumar2022} for analysis demanding repeated system solves like uncertainty quantification, sensitivity analysis and parametric investigations.

\begin{table}[htb]
	\caption{Specifications and summary of \ac{MOR} results for the three aircraft models}
	\label{tab:MOR_ModelOverview}
	\resizebox{\textwidth}{!}{\begin{tabular}{p{1.5cm}p{2.0cm}p{2.2cm}p{2.4cm}p{1.5cm}p{1.8cm}ll}
		\hline
		Models                   & Frequency range         & l-fROM window  & Full/reduced dimension & Number of excited \ac{DoF}s & Number of extracted eigenmodes & Error norm$^1$        & Speedup$^{2}$ \\ \hline
		Model 1                  & {[}10,258{]} Hz                    & {[}10,258{]} Hz                    & 339265/490        & 37995                  & 361                  & $5.51\cdot10^{-3}$ &  718       \\ \hline
		\multirow{2}{*}{Model 2} & \multirow{2}{*}{{[}258,578{]} Hz}  & {[}260,418{]} Hz                    & 632967/630        & 72963                  & 333                  & $8.21\cdot10^{-3}$         & 1304         \\
		&                                         & {[}418,578{]} Hz                    & 632967/610        & 72963                  & 370                  & $5.03\cdot10^{-3}$         & 1411        \\ \hline
		\multirow{3}{*}{Model 3} & \multirow{3}{*}{{[}578,1000{]} Hz} & {[}578,718{]} Hz                    & 1080621/570          & 116337                 & 399                    & $5.81\cdot10^{-3}$                 & 3488       \\
		&                                         & {[}718,858{]} Hz                & 1080621/540          & 116337                 & 423                    & $2.64\cdot10^{-3}$                 & 3287       \\
		&                                         & {[}858,1000{]} Hz                  & 1080621/550          & 116337                 & 474                    & $6.59\cdot10^{-3}$                 & 2889      \\ \hline
	\end{tabular}}
	$^1$ Maximum relative error norm over frequency, $\epsilon_\mathrm{max}$ \\
	$^2$ Speedup per frequency solve for the dense solver time in \ac{MOR} online phase for ROM computation with respect to the sparse solver time for the direct solution from Tab.~\ref{tab:finalMeshes}
\end{table}

Convergence of the computed \ac{fROM} is ensured by complying with the desired relative error tolerance of $1\cdot10^{-2}$. To ensure the convergence or the quality of the generated \ac{fROM} in every greedy iteration within an adaptive framework, we use the expensive classical relative error measure $\mathbf{\epsilon}(\omega) = \left| \mathbf{y}(\omega) - \mathbf{y}_R(\omega) \right|/\left| \mathbf{y}(\omega) \right|$ and the maximum relative error measure $\epsilon_\mathrm{max} = \arg \max(\mathbf{\epsilon}(\omega))$ to evaluate the fROM quality for the entire frequency domain. This can be replaced with cheaper estimates, which enable the true advantage of performing \ac{MOR}. 

The modal density of the considered aircraft model spanning the desired frequency interval was challenging requiring more expansion points to generate an accurate fROM. This leads to an expensive \ac{MOR} offline phase owing to the time required for factorisation of the full-order model matrices per expansion point. For the computation of projection basis as in Eqn.~\eqref{eq:FOKSM}, a direct LU factorisation using the MUMPS solver is performed as detailed in Sec.~\ref{sec:problem}. For higher frequencies and large-scale matrices, deploying iterative solvers can be significant to accelerate the \ac{MOR} offline process. The \ac{DD} approach, presented in Sec.~\ref{sec:DD}, is also interesting in combination with recycling strategies \cite{Ahuja2015} for faster convergence of iterative solutions while computing moments by solving a linear system of equations with multiple right-hand sides.

In summary, the presented \ac{MOR} approaches deliver accurate reduced order models approximating the considered aircraft model response with the desired accuracy. A significant reduction in dimension and achieved speedup in computational time per frequency solve are observed. Compared to the reference solver performance, the total computational time is reduced from 42.1 hours (or 2.6 hours with \ac{DD} setting in Sec.~\ref{sec:DD}) to just 4.26 seconds in the online phase for considered reduction setting. The time required for the offline phase is therefore considered a one-time cost. Also, the influence of frequency-dependent loading and material parameters on the system's response is approximated well by the generated \acp{fROM}. We thereby highlight the potential advantages of using surrogates obtained with \ac{rA-Krylov} for practical large-scale problems, but at the same time demanding considerable improvement in methods used to overcome various challenges described in this section.

\section{SUMMARY, CONCLUSION AND OUTLOOK}
\label{sec:summary}

The contribution examined the presented three efficient solving strategies (frequency- and domain-adaptive meshes, \ac{DD}, \ac{MOR}) applied to the vibroacoustic FE model of an aircraft fuselage segment with the objective to perform cabin noise simulations. We compare the speed-up and memory-saving of the three approaches to a direct solution using the massively parallelised direct solver MUMPS on a high-end workstation. The conventional workflow demand, per frequency solve for the finest mesh configuration, yields $300.3$ seconds accounting for a total of $42.1$ hours and requires $164.3$ GB of maximum memory.

Foremost, the frequency- and domain-adaptive discretisation relieves computational expenses by enabling the generation of relatively coarser FE meshes allowing more than $50\%$ reduction in the system dimensions without actually compromising the accuracy of the solution. A performance gain of $95\%$ in computation time is obtained, yielding the full solution in $2.8\,$hours and $29.2\,$GB of resident memory in RAM. The significant decrease in computational costs depicts an indispensable standard for cabin noise simulations. For even higher frequency ranges, at least at the coincidence frequency the meshes are again conform. This fact relativises the performance gain with increasing frequency and will converge to the advantage of frequency-dependent meshes. \ac{DD} and \ac{MOR} techniques are studied by using the proposed optimally chosen meshes. 

\ac{DD} techniques enable a flexible framework to perform iterative solving that yields converging solutions for the presented fuselage model with challenging heterogeneous domains. Physical-domain based treatment and application of suitable preconditioners is the key to handle systems of such complexity that diminishes the chance of solving with iterative solvers. Our investigations yield significantly faster and feasible solves using GMRES and GASM preconditioning with overlapping subdomains. The resulting computational effort now entails $2.6$ hours using a maximum memory of only $4.5$ GB. Hence, we observe the advantage of performing computations at reduced memory requirements for large-scale models.

\ac{MOR} using \ac{rA-Krylov} provides a fast-track alternative route to compute the desired solution instantaneously from the generated reduced models. As the expensive large-scale sparse matrices are now replaced with significantly smaller dense matrices (around 600 \ac{DoF}s for our models), the solving procedure delivers a system response within a few milliseconds and negligible memory usage for all aircraft meshes. The offline phase is observed to be expensive due to the requirement of comparatively more number of expansion points so as to capture the dynamics of the aircraft system sufficiently. Regardless, one is obtained with reduced order model matrices of very small dimensions that can serve as the surrogate basis to perform faster computations for investigations involving repeated system solve like uncertainty quantification, sensitivity analysis and parametric studies. As a result, the computational effort for the whole simulation run is now reduced to just $4.3$ seconds in the \ac{MOR} online phase at least memory usage. However, we underline the used \ac{MOR} setting that may vary for different applications.

In summary, all three approaches support efficient computations and serve as strategies to accelerate expensive processes such as product design and optimisations. We identify more challenges and potential ideas for future research. All three approaches can benefit from iterative solving techniques in the parallel computing setting by exploiting the vibroacoustic nature and physicality. Also, repeated system solves can utilise recycling techniques to further accelerate the solving phase.

Future and ongoing research can focus on the utilisation of physics-based preconditioners to increase the efficiency of iterative solvers in vibroacoustics, while load balancing in parallelisation plays an important role in distributing the physical subdomains equally among processes. 
In the scope of \ac{MOR} techniques, the focus can be laid on a more efficient offline phase that deals with the transport of domain-based preconditioning to construct stable reduced order models for coupled systems. Efficient handling of similar highly dynamic systems in context of surrogate modelling is seen as important. Further research projects also include the study of resulting uncertainties propagating in the simulation and further on in the surrogate modelling process.

\section*{ACKNOWLEDGEMENTS}

We would like to acknowledge the funding by the Deutsche Forschungsgemeinschaft (DFG, German Research Foundation) under Germany's Excellence Strategy – EXC 2163/1 - Sustainable and Energy Efficient Aviation – Project-ID 390881007.

\bibliography{literature}

\end{document}